\documentclass[final,5p,times,twocolumn]{elsarticle}
\usepackage{amssymb}
\usepackage{lipsum}
\usepackage{upgreek,xspace}
\usepackage[colorlinks=true, allcolors=blue]{hyperref}
\usepackage{multirow}
\usepackage{microtype}
\usepackage{tabularx}
\usepackage{booktabs}
\usepackage[table]{xcolor}
\usepackage{array}
\usepackage{rotating}
\usepackage{lineno}

\journal{Nuclear Physics A}

\begin{document}

\begin{frontmatter}

\title{Development and Characterization of MPGD-based Transition Radiation
Detectors}

 \author[first]{Lauren Kasper\corref{cor1}}
 \ead{lauren.n.kasper@vanderbilt.edu}
 \affiliation[first]{organization={Vanderbilt University},
            addressline={Department of Physics and Astronomy}, 
            city={Nashville},
            postcode={37235}, 
            state={TN},
            country={US}}
 \author[second]{Alexander Austregesilo}
 \affiliation[second]{organization={Thomas Jefferson National Accelerator Facility},
             city={Newport News},
             postcode={23606},
             state={VA},
             country={US}}
\author[second]{Fernando Barbosa}
\author[second]{Cody Dickover}
\author[second]{Sergey Furletov}
\author[second]{Yulia Furletova}
\author[second]{Kondo Gnanvo}
\author[first]{Senta Vicki Greene}
\author[second]{Lubomir Pentchev}
\author[second]{Sourav Tarafdar}
\author[first]{Julia Velkovska}

\cortext[cor1]{Corresponding author}

\begin{abstract}

Transition Radiation Detectors (TRDs) are useful for electron identification and hadron suppression in high energy nuclear and particle physics experiments. Conventional wire-chamber TRDs face operational limitations due to space charge effects, motivating the replacement of the amplification stage with MicroPattern Gaseous Detectors (MPGDs). This work explores different MPGD technologies -- Gas Electron Multiplier (GEM), Micro-Mesh Gaseous Structure (Micromegas), and Resistive Micro-Well ($\upmu$RWELL) -- as alternative TRD amplification stages. We report on the design, construction, and in-beam characterization of multiple MPGD-based TRD prototypes exposed to 3--20\,GeV mixed electron–hadron beams at the Fermilab Test Beam Facility and at the CERN SPS H8 beamline. Each detector consisted of a multi-layered radiator, an approximately 2\,cm deep drift region, an MPGD amplification stage optimized for X-ray transition radiation detection in a Xe:CO$_{2}$ (90:10) gas mixture, and a two-dimensional readout. The GEM-based TRD prototype achieved a pion suppression factor of about 8 at 90\% electron efficiency, while the Micromegas-based prototype -- with an added GEM preamplification layer -- demonstrated improved operational stability and clear TR photon discrimination. The $\upmu$RWELL prototype achieved stable operation but limited signal gain.\texttt{Geant4}-based studies confirmed the observed trends and highlighted the sensitivity of the TR yield to cathode material and radiator configuration. These studies represent the first in-beam measurements of Micromegas- and $\upmu$RWELL-based TRDs, along with discussion of the performance capabilities of a triple-GEM-TRD. The results demonstrate the feasibility of MPGDs as scalable, high-rate amplification structures for next-generation TRD applications.

\end{abstract}


\end{frontmatter}

\section{Introduction}
\label{introduction}

Tracking and identifying charged particles are significant components of most high energy nuclear and particle (HENP) physics experiments. Precise electron identification grants access to processes like quarkonia and heavy flavor production where leptonic decay channels are among the cleanest probes. It is therefore useful to suppress overwhelming hadronic backgrounds for these channels and provide reliable signatures for rare and heavy states.

Transition Radiation Detectors (TRDs) are well suited for this task due to their ability to separate electrons from hadrons over a broad momentum range ($\sim$1--150\,GeV/$c$)~\cite{PDG}. 
 In this momentum range, only electrons produce X-ray transition radiation (TR) in appropriate radiators, suitable for detection in heavy gases in the 3--40\,keV energy region. Due to its low material budget, TRD can be combined with other electron identification methods, such as calorimetry, for improved electron/hadron separation and cross-calibration of the two detector systems. Both regular foil radiators and irregular fleece radiators are commonly used, each offering tradeoffs in photon yield, spectral distribution, and material budget.

A typical gaseous TRD consists of a radiator, gas volume, and amplification stage to detect the ionization deposited in the gas. The ALICE TRD~\cite{AliceTRD} is one example of a modern TRD with typical drift chamber amplification. Space charge effects in traditional wire-based amplification structures have been previously reported~\cite{ALICESpaceCharge}~\cite{JLabWireChambers}. While able to be mitigated, these effects have resulted in performance issues for TRDs that have limited their application in HENP experiments. This motivates the replacement of the amplification stage with MicroPattern Gaseous Detector (MPGD) technologies such as Gas Electron Multiplier (GEM)~\cite{GEM}, Micro-Mesh Gaseous Structure (Micromegas)~\cite{MMG}, and Resistive Micro-Well ($\upmu$RWELL)~\cite{URW}. These technologies offer excellent rate capability, reduced ion backflow, improved energy resolution, and simplified scalability to large systems.

Previous efforts at Thomas Jefferson National Accelerator Facility (Jefferson Lab) have demonstrated a proof-of-principle GEM-based TRD prototype, achieving an electron–pion (e/$\pi$) rejection factor greater than 5 at 90\% electron efficiency, along with three-dimensional charged-particle tracking capability~\cite{NIMGEMTRD1}. In this paper, we build on this previous work to explore Micromegas and $\upmu$RWELL as amplification choices in a TRD application for the first time. Both MPGD options possess practical advantages such as lower operating voltages, fewer high-voltage channels, more uniform gain over larger surface area, and simplified construction compared to GEMs, while retaining comparable energy resolution.

We present the design, development, and beam-test characterization of several MPGD-based TRD prototypes. In Section~\ref{design}, we introduce the concept of triple-GEM-, $\upmu$RWELL-, and Micromegas-based TRDs and describe the design and construction of prototypes. Section~\ref{beam_tests} describes the in-beam studies of these prototypes in mixed hadron--electron beams at the Fermilab Test Beam Facility (FTBF) and the CERN SPS H8 beamline. Section~\ref{results} presents findings on operational stability, detector efficiency, timing performance, and e/$\pi$ separation while Section~\ref{discussion} discusses the implications of these results and complementary \texttt{Geant4}-based simulations. Section~\ref{conclusion} summarizes our conclusions on these different MPGD technologies as amplification layers in a TRD concept and outlines future tests planned for these novel detectors.

\section{Prototype Development and Design}
\label{design}

Optimizing TRD performance requires careful consideration of several detector parameters: the entrance window material and thickness to maximize transmission of soft X-rays without compromising stability of operation, the drift region length to absorb higher-energy TR photons, and the radiator design to maximize TR production while minimizing self-absorption. The design concept for the TRD prototypes developed in this study was guided by the goal of simultaneously maximizing the probability of detecting TR photons while minimizing material budget and dead area, within the practical constraints of iterative R\&D and prototyping. Each detector consists of a radiator upstream of a gaseous drift region followed by an MPGD amplification stage and a two-dimensional strip readout plane. The drift region, ranging from 20--28\,mm in depth, was chosen to efficiently absorb TR photons in the 3--40\,keV range while maintaining stable electric field operation; previous MC simulations showed that expansion beyond this range does not significantly improve pion suppression performance~\cite{NIMGEMTRD1}. Cathode foils were selected with consideration for balancing soft X-ray absorption without compromising overall mechanical and electrical robustness. Both fleece-type and foil-type radiators were tested, enabling a direct comparison of irregular versus regular radiator geometries for TR generation. The signal amplification stage was realized with the following MPGD technologies: GEMs, Micromegas, and $\upmu$RWELLs. Table~\ref{tab:detector_params} displays a comparison of major design parameters between each of the tested prototypes.

\begin{table*}[!h]
    \centering
    \scriptsize
    \renewcommand{\arraystretch}{1.2}
    \begin{tabular}{llllll}
\hline
Parameter&GEM--TRD&GEM--TRD&Micromegas&Micromegas&$\upmu$RWell\\
&\_v1&\_v2&--TRD&+GEM--TRD&--TRD\\
\hline
Window material&Kapton&Kapton&Mylar&Kapton&Kapton\\
Window thickness [$\upmu$m]&25&55&25&55&50\\
Dead gas gap [$\upmu$m]&400&0&1000&0&0\\
Cathode material&Cr&Cu&Stainless Steel&Cu&Cr\\
Cathode thickness [$\upmu$m]&0.2&5&18&5&0.2\\
Drift gap depth [mm]&21&19.9&28&25.5&25\\
Chamber depth [mm]&28&26.5&29.5&28.5&25.5\\
\hline
\end{tabular}
    \caption{Main design parameters of the various MPGD-based TRD prototypes developed and tested.}
    \label{tab:detector_params}
\end{table*}

\subsection{Triple-GEM-TRD}

\begin{figure}[!h]
\centering
\frame{\includegraphics[width=0.89\columnwidth,trim={750pt 1050pt 1250pt 1150pt},clip]{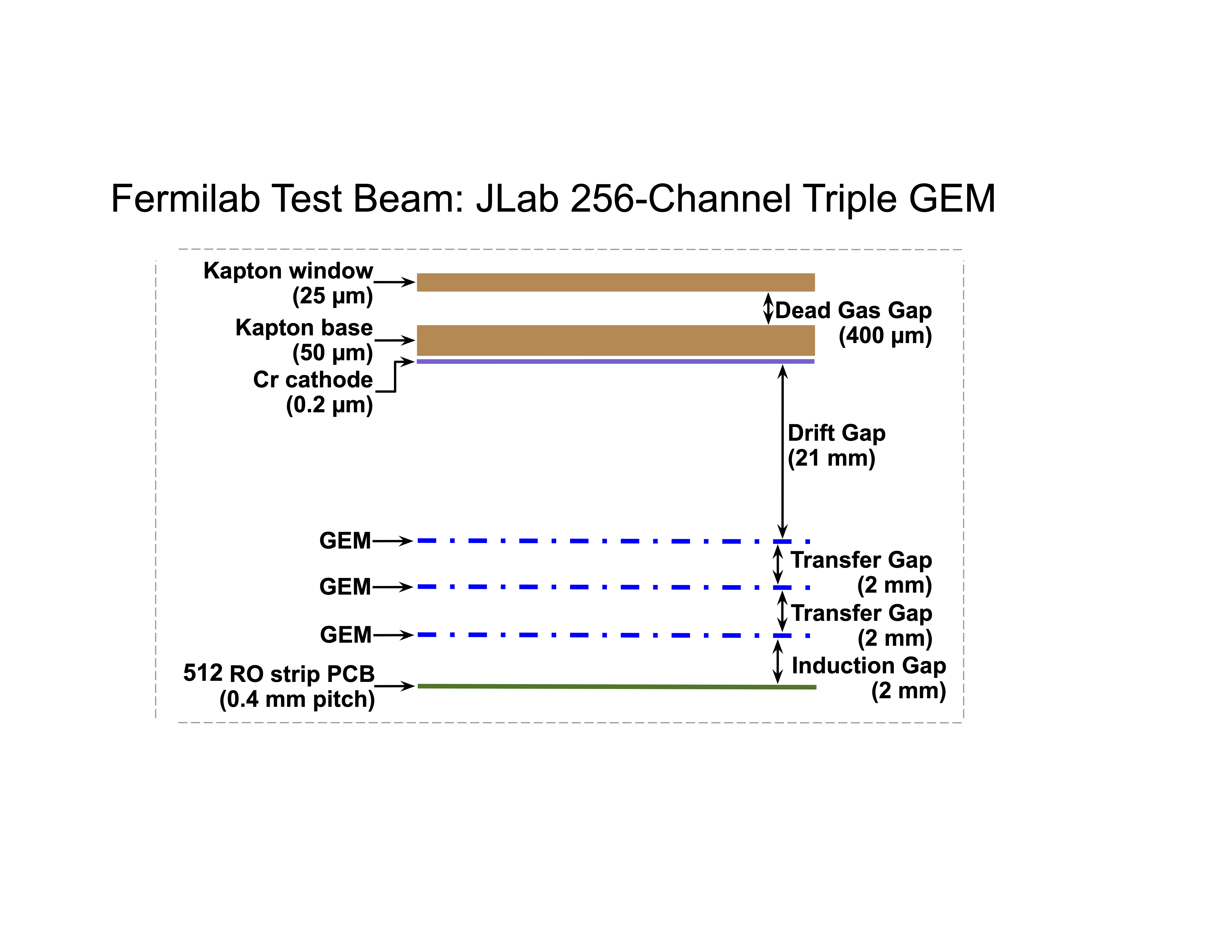}}
\frame{\includegraphics[width=0.89\columnwidth,trim={750pt 1050pt 1250pt 1150pt},clip]{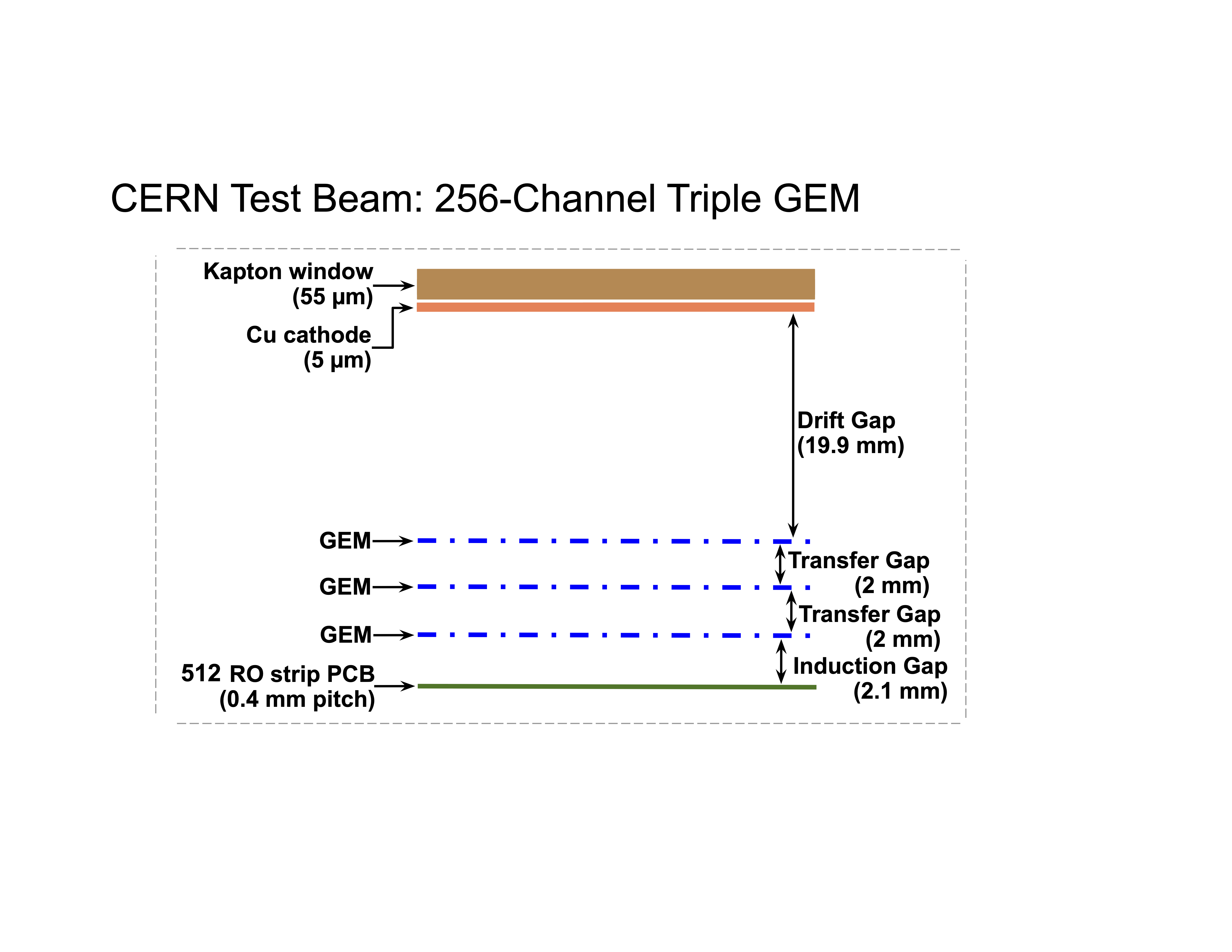}}
\caption{\label{fig:gemtrd} (\textit{Top}) Cross-sectional schematic of the GEM-TRD\_v1 design, as tested at FTBF; (\textit{Bottom}) cross-sectional schematic of the GEM-TRD\_v2 design, as tested at CERN SPS. Note the schematics are not to scale.}
\end{figure}

The GEM-TRD serves as the reference technology for these tests and was previously studied in detail at Jefferson Lab~\cite{NIMGEMTRD1}. The detector employs a standard $10~\times~10$\,cm$^2$ triple-GEM amplification stack with transfer and induction gaps of 2\,mm.
Charge readout is provided by a two-dimensional orthogonal X–Y strip layout comprising of 256 strips per dimension with a pitch of 400\,$\upmu$m. 
Two separate GEM-TRD prototype designs were evaluated. The first, referred to as GEM-TRD\_v1, retained the configuration described in~\cite{NIMGEMTRD1} originally tested at Jefferson Lab. GEM-TRD\_v1 was tested to extend the original proof-of-principle studies to hadron beams at FTBF. While GEM-TRD\_v1 successfully demonstrated TRD operation, its design incorporates a $\sim$400\,$\upmu$m dead gas gap immediately following the entrance window.
If scaled to larger detector areas, this dead region is expected to increase in depth due to entrance-window bulging under operating gas pressure, leading to enhanced absorption of soft transition radiation photons in the heavy gas before reaching the active drift region.
To isolate and eliminate this effect, a second triple-GEM prototype, referred to as GEM-TRD\_v2, was constructed without an entrance-window-induced dead gas gap. In this design, the 25\,$\upmu$m Kapton entrance window present in GEM-TRD\_v1 was removed entirely and a 5\,$\upmu$m copper cathode foil supported on 55\,$\upmu$m Kapton was installed. The choice of 5\,$\upmu$m Cu reflects a more conventional choice for cathode material and avoids potential mechanical and electrical stability concerns associated with the ultra-thin 200\,nm Cr cathode employed in the GEM-TRD\_v1 design.
Schematics of these two constructions are shown in Figure~\ref{fig:gemtrd}. For both designs of the GEM-TRD, a voltage divider was used for the triple-GEM amplification structure as a means of reducing the necessary number of HV channels from seven to two. Nominal voltage settings and electric field values are provided in~\ref{App_a}.

\subsection{Micromegas-Based TRD}

\begin{figure}[!h]
\centering
\frame{\includegraphics[width=0.89\columnwidth,trim={750pt 1000pt 1300pt 1150pt},clip]{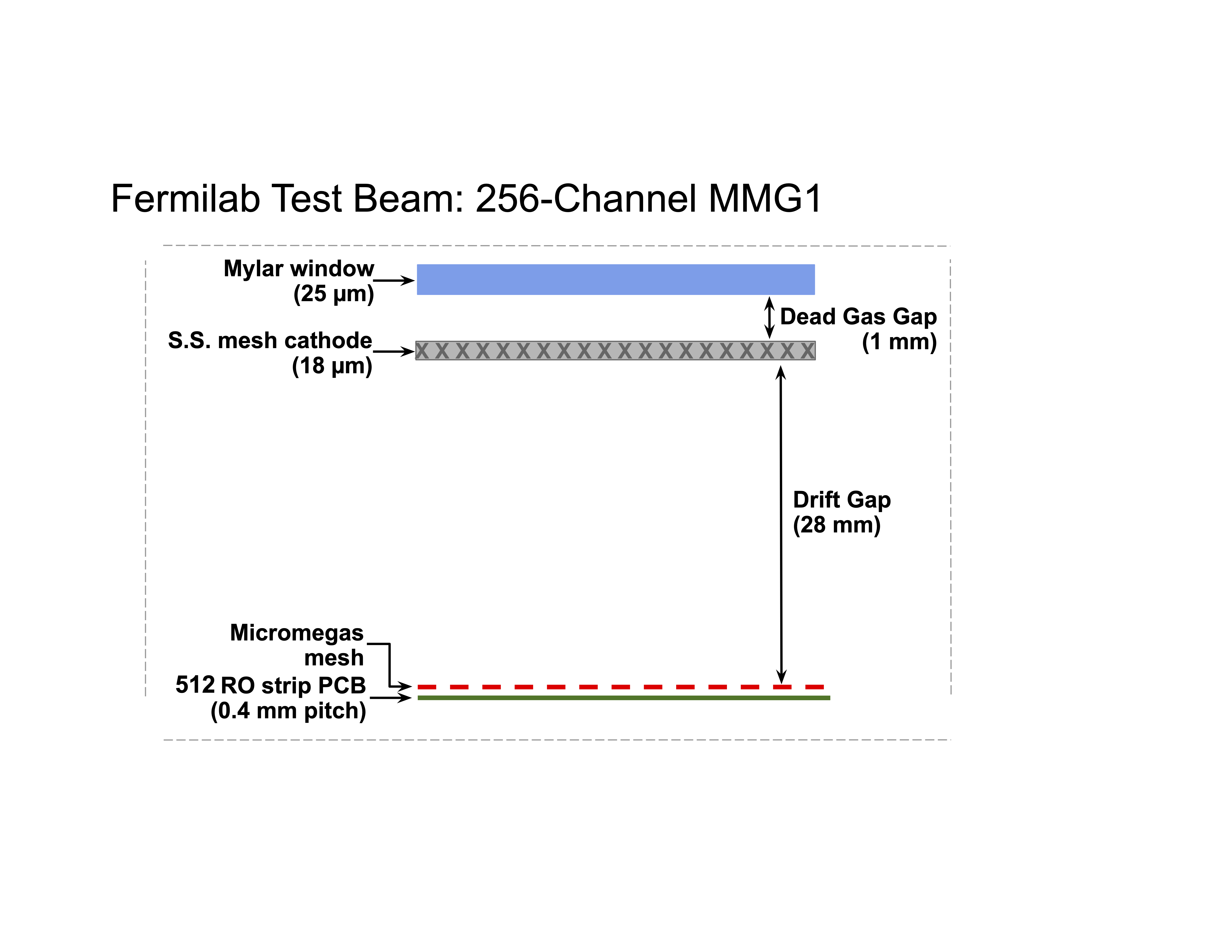}}
\frame{\includegraphics[width=0.89\columnwidth,trim={750pt 1150pt 1300pt 1175pt},clip]{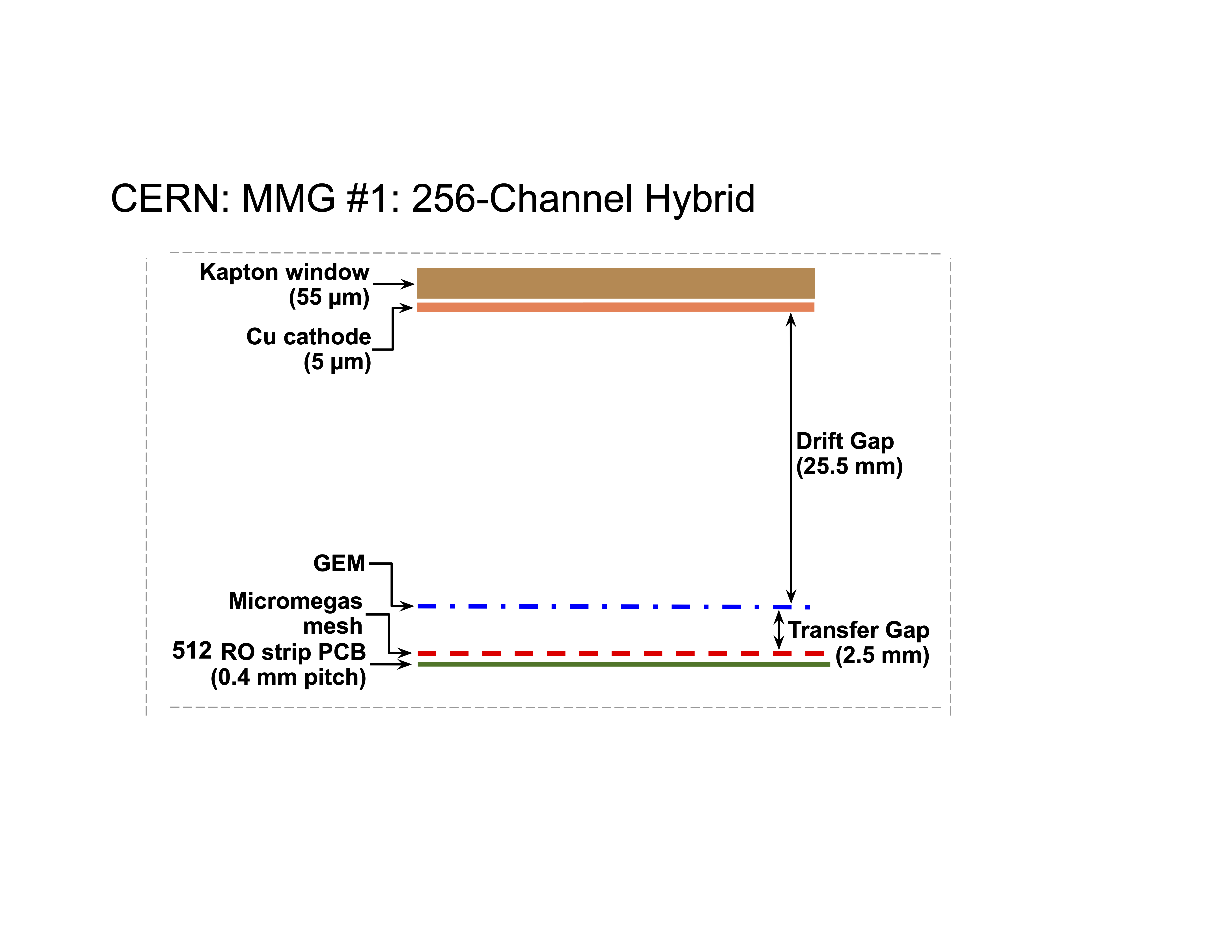}}
\caption{\label{fig:mmgtrds} (\textit{Top}) Cross-sectional schematic of the Micromegas-TRD prototype, as tested at FTBF; (\textit{Bottom}) cross-sectional schematic of the Micromegas+GEM-TRD, as tested at CERN SPS. Note the schematics are not to scale.}
\end{figure}

A Micromegas-TRD prototype was constructed at Vanderbilt University and tested in-beam at FTBF. The readout board is of the same layout and design as the GEM-TRD, employing a two-dimensional orthogonal X–Y strip layout with 256 strips per dimension at 400\,$\upmu$m pitch. The drift region was 28\,mm, and the cathode was a fine stainless steel mesh of 18\,$\upmu$m thickness mounted approximately 1\,mm below a 25\,$\upmu$m Mylar entrance window. Following the FTBF tests, the prototype design was modified by introducing a single GEM foil upstream of the Micromegas mesh, separated by a 2.5\,mm transfer gap and thereby reducing the drift gap to 25.5\,mm. The GEM preamplification stage was introduced to improve the overall signal gain and reduce the probability of discharges, which caused limitations in stable operation during the FTBF tests. The introduction of a GEM preamplification layer also lead to the use of a voltage divider for the amplification structure to reduce the number of input HV channels to two. Nominal voltage settings and electric field values are given in~\ref{App_a}. The Micromegas+GEM-TRD also utilized an entrance window of a 5\,$\upmu$m Cu foil on 55\,$\upmu$m Kapton, for direct comparability to the GEM-TRD\_v2 design and in order to remove the non-sensitive gas gap present in the Micromegas-TRD configuration. Figure~\ref{fig:mmgtrds} displays a cross section of both Micromegas-based prototype constructions.

\subsection{\textmu RWELL-TRD}

\begin{figure}[!h]
\centering
\frame{\includegraphics[width=0.95\columnwidth,trim={25pt 5pt 25pt 5pt},clip]{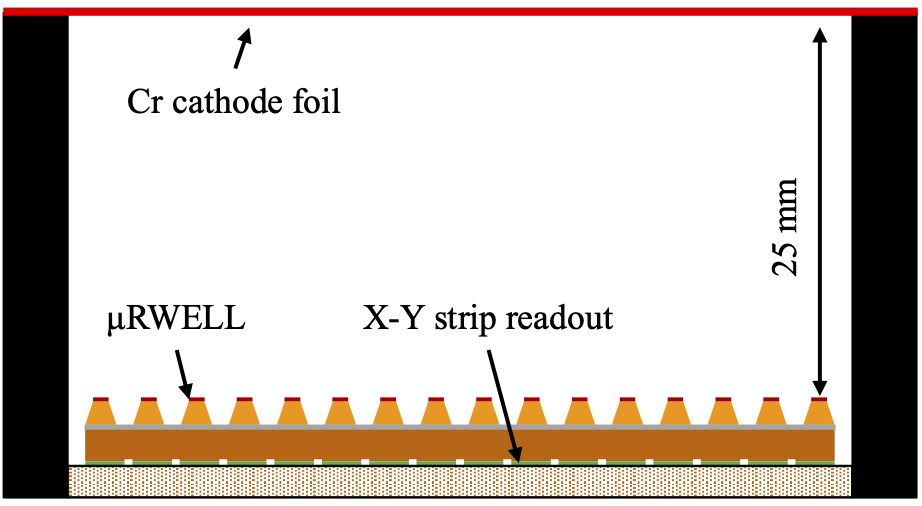}}
\caption{\label{fig:urwelltrd} Cross-sectional schematic (not to scale) of the $\upmu$RWELL-TRD prototype.}
\end{figure}

A $\upmu$RWELL-TRD prototype was assembled and tested at Jefferson Lab prior to in-beam tests. Its distinguishing feature is the use of a chromium-coated polyimide cathode foil (200\,nm Cr on 50\,$\upmu$m Kapton), reducing the material budget seen by TR photons while maintaining operational robustness. The drift gap was 25\,mm. The amplification stage consisted of a resistive Micro-Well structure coupled to an X–Y strip layout with 128 strips per dimension, equating to a 800\,$\upmu$m strip pitch as described in~\cite{capaSh_urwell2022}. Figure~\ref{fig:urwelltrd} illustrates the detector construction. 

\subsection{Readout Electronics and DAQ System}

Fast signal collection is essential for TRDs in order to separate clusters along the track and distinguish the TR photon-induced signals from energy deposited in the gaseous drift region by all charged particles. For these prototypes, flash ADC modules (fADC-125)~\cite{fadc} with 8\,ns sampling time developed by Jefferson Lab were used, and signals were amplified with fast (10\,ns peaking time) GAS-II preamplifiers~\cite{GAS2} before digitization. These fast front-end electronics require higher detector gas gain than commonly used APV25-based readout~\cite{apv25}, but are critical for characterizing the timing and amplitude distributions of TR photon clusters. Regarding the readout electronics setup, 240 out of 256 strips on the X-plane of each detector are connected individually to ADC modules and read out in parallel, while all 256 strips of the Y-plane are read out serially with the SRS.

\section{In-Beam Tests}
\label{beam_tests}

The various detectors described in the previous section were tested in mixed hadron--electron beams at FTBF and CERN SPS. Neither the Micromegas-TRD nor the $\upmu$RWELL-TRD could reach sufficient signal gain to enable a decisive measurement of hadron suppression. This observed signal gain limitation of a single amplification layer motivated the modification of the Micromegas prototype into a hybrid MPGD design, where a GEM preamplification stage was implemented. The Micromegas+GEM-TRD also utilized an entrance window of a 5\,$\upmu$m Cu foil adhered to 55\,$\upmu$m Kapton, for direct comparability to the GEM-TRD\_v2 design as discussed in Section~\ref{design}.


\begin{figure}[!h]
\begin{center}
\includegraphics[width=0.99\columnwidth,trim={5pt 12pt 5pt 5pt},clip]{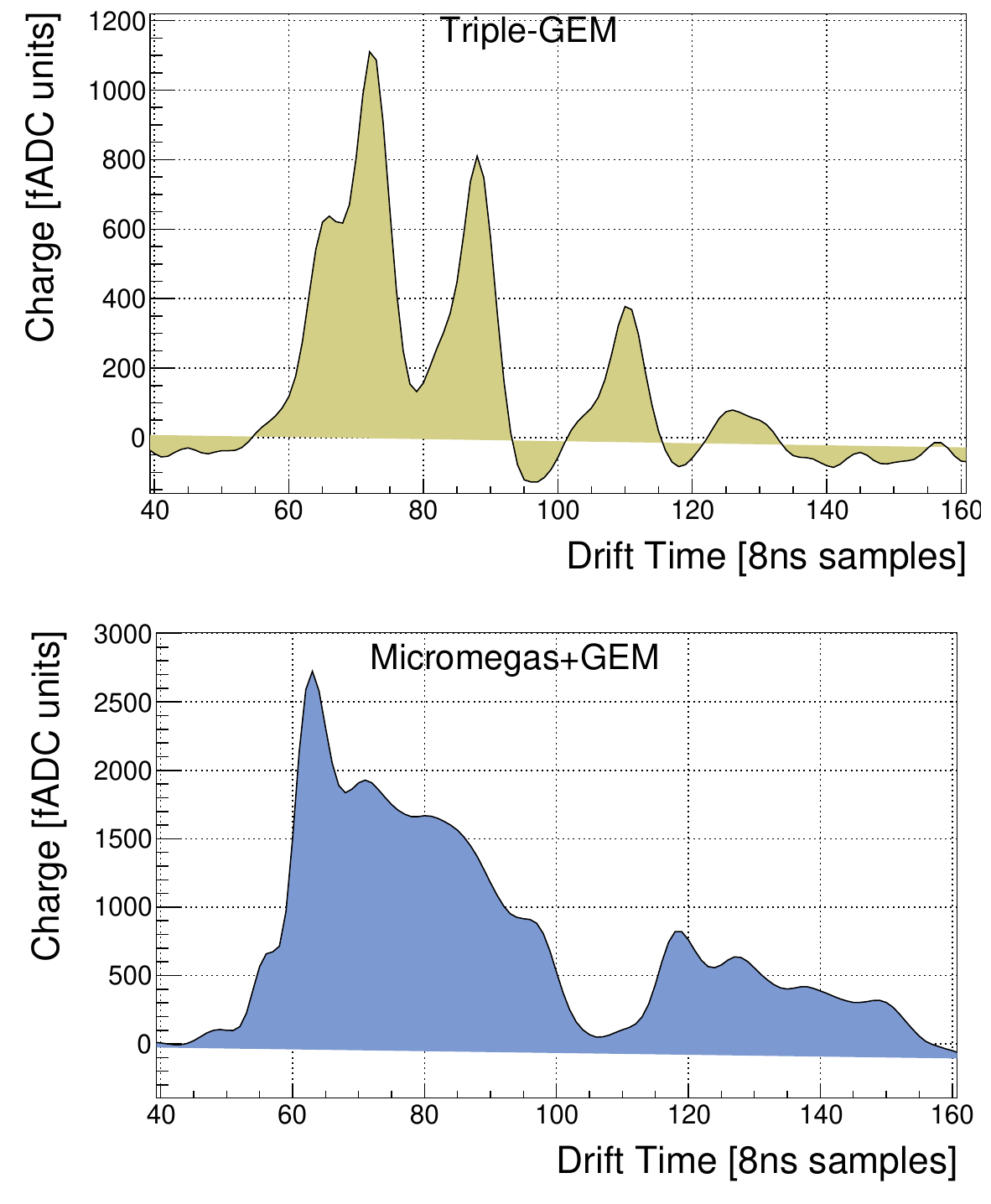}
\caption{Example of a typical raw fADC-125 waveform from an electron signal for the GEM-TRD\_v2 (\textit{Top}) and the Micromegas+GEM-TRD (\textit{Bottom}).}
\label{fig:raw_f125}
\end{center}
\end{figure}

\subsection{FTBF Experimental Setup}
\label{ftbf_setup}
\begin{figure}[!h]
\begin{center}
\includegraphics[width=\columnwidth,trim={0pt 25pt 25pt 65pt},clip]{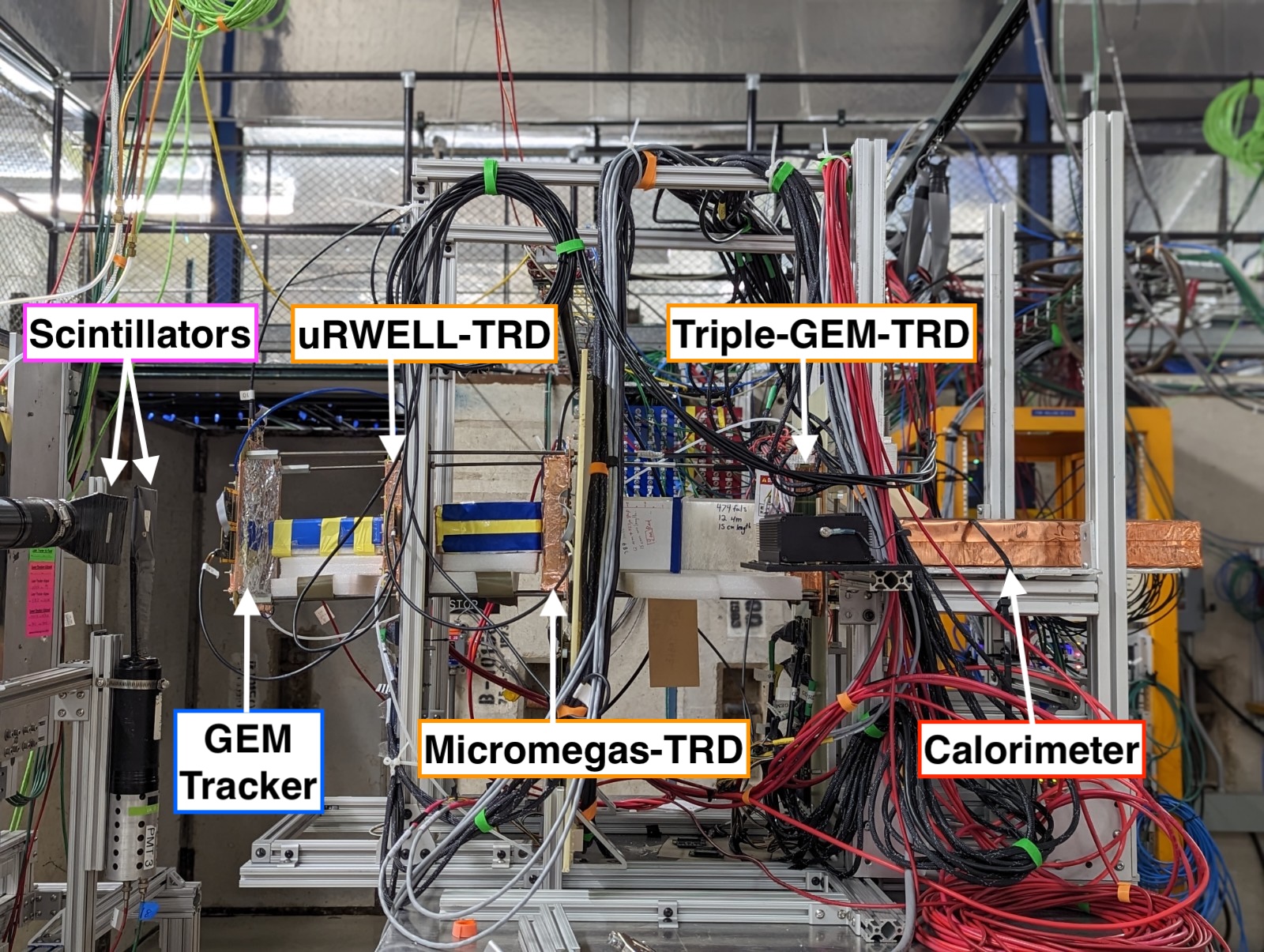}
\includegraphics[width=\columnwidth,trim={48pt 240pt 80pt 110pt},clip]{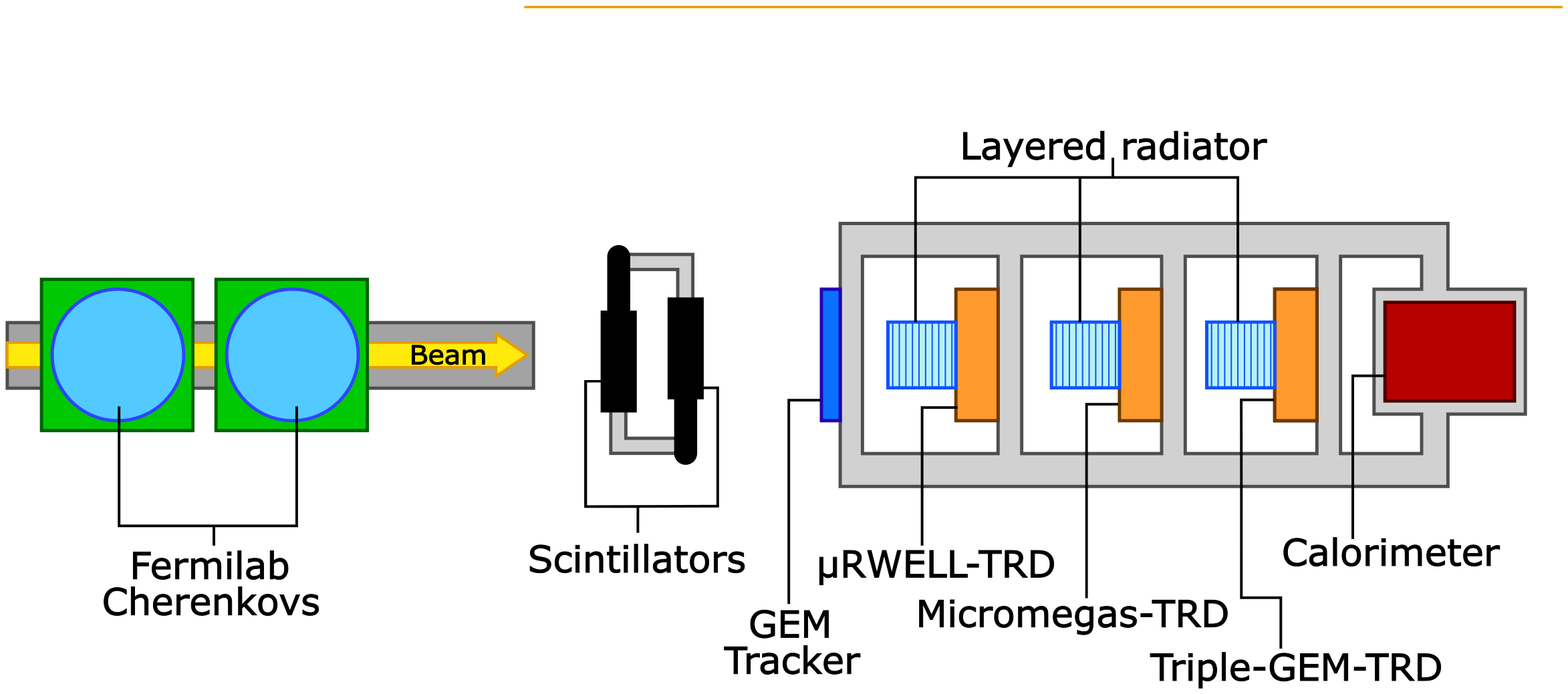}
\caption{(\textit{Top}) Experimental setup used for the FTBF test beam measurements. Note that the Fermilab-owned Cherenkov detectors are not pictured, since they are upstream from the setup. (\textit{Bottom}) Schematic of the detector layout for the FTBF test beam experimental setup (not to scale). The various components are described in the document text.}
\label{fig:Fermi_Setup}
\end{center}
\end{figure}

In May 2023, the GEM-TRD\_v1, Micromegas-TRD, and $\upmu$RWELL-TRD prototypes were tested at FTBF in mixed electron–hadron beams of 3\,GeV and 10\,GeV. Electron fractions were approximately 70\% at 3\,GeV and 50\% at 10\,GeV. A standard triple-GEM tracker with better than $80\,\upmu$m spatial resolution was placed upstream of the setup in order to provide a track reference. It utilized APV25 front-end ASICs to shape the signals from all electrode channels, which were subsequently digitized using a Scalable Readout System (SRS)~\cite{srs}. Triggering was accomplished by two scintillators overlapped in front of the active area of the detectors. The TRD prototypes were operated in a Xe:CO$_{2}$ (90:10) gas mixture. Cherenkov detectors in the beamline were tuned to discriminate electrons from hadrons and used for external electron and pion sample selection. Additional particle identification (PID) was provided by a 7-cell lead--tungsten electromagnetic calorimeter supplied by Jefferson Lab. Figure~\ref{fig:Fermi_Setup} shows the experimental setup. The calorimeter provides a pion suppression factor of roughly 100, leading to sample purity of about 99\%; the Cherenkov detectors provide an additional redundancy in external PID, resulting in adequate sample purity.

Two radiator types were used in order to perform tests comparing the TR-yield of different materials at FTBF: fleece radiator material with irregular structure from the HERMES and ZEUS experiments~\cite{hermes}~\cite{zeus} and radiators constructed with regularly spaced Mylar foils having 12.5\,$\upmu$m thickness separated by netted spacers of $\sim$300\,$\upmu$m thickness. The fleece material had a density of roughly 0.09\,g/cm$^{3}$, and the Mylar foil radiators had a density of approximately 0.065\,g/cm$^{3}$. Table~\ref{tab:rad_layouts} describes the radiator parameters.

\begin{figure}[!h]
\begin{center}
\includegraphics[width=\columnwidth,trim={0pt 20pt 5pt 0pt},clip]{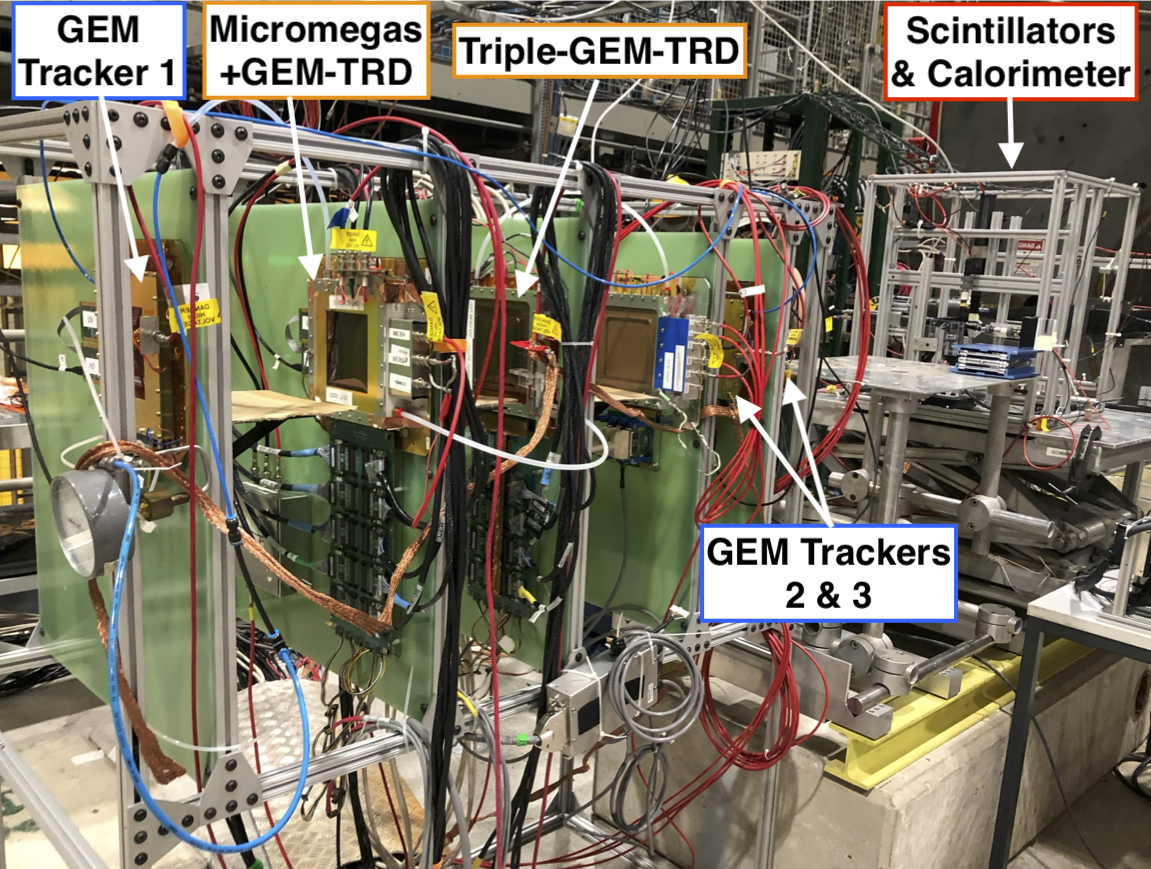}
\includegraphics[width=\columnwidth,trim={2pt 210pt 40pt 120pt},clip]{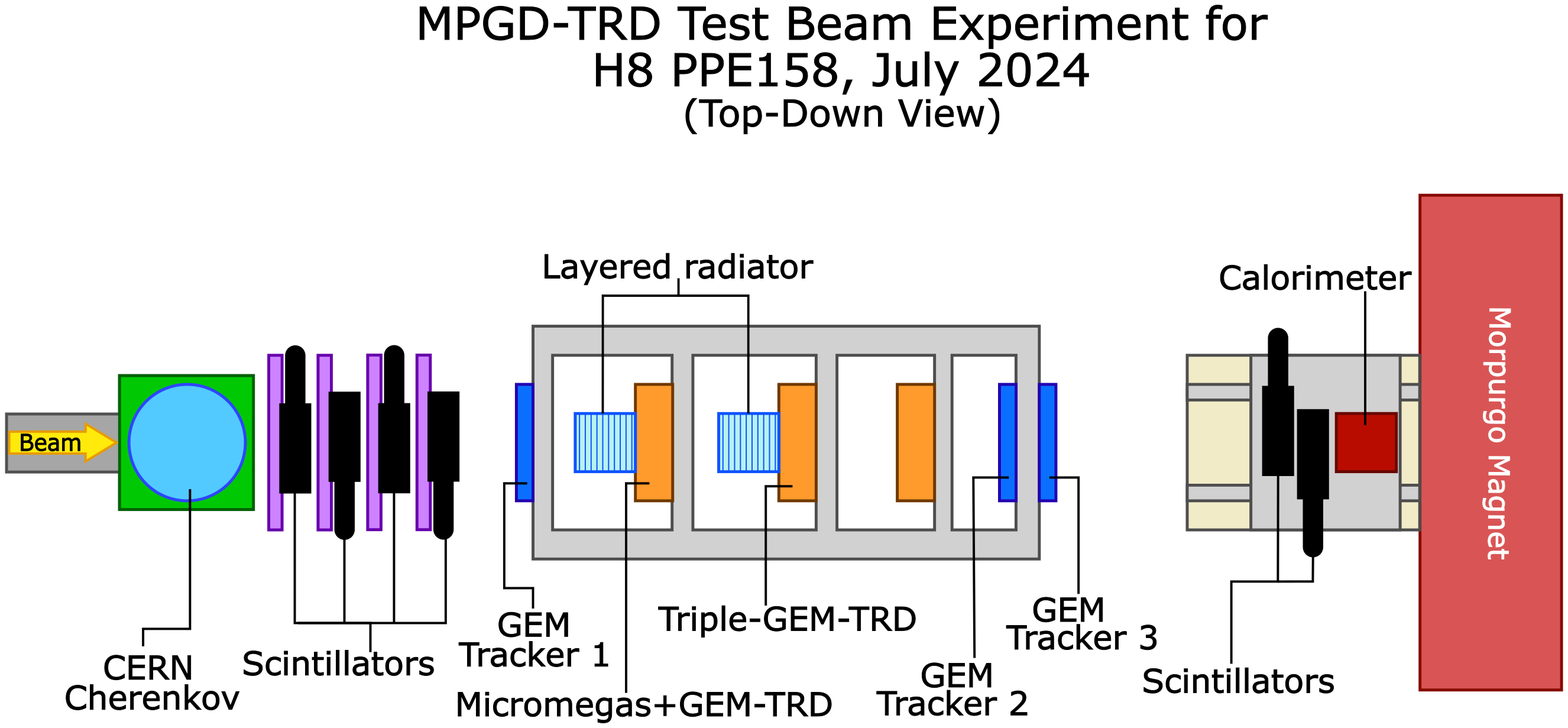}
\caption{(\textit{Top}) Experimental setup used for the CERN SPS test beam measurements --- note that the scintillators and CERN-housed Cherenkov detector are not pictured since they are upstream from the setup. (\textit{Bottom}) Schematic of the detector layout for the CERN test beam experimental setup (not to scale). The various components are described in the document text.}
\label{fig:CERN_Setup}
\end{center}
\end{figure} 

\subsection{CERN SPS Experimental Setup}
\label{cern_setup}

In July 2024, tests of the GEM-TRD\_v2 and Micromegas+GEM-TRD were carried out at CERN SPS H8 beam line with 20\,GeV mixed electron–hadron beams. The setup included three standard triple-GEMs for precise tracking, an upstream Cherenkov detector for PID, upstream scintillators for timing and triggering, and a downstream single-cell Pb-glass calorimeter for additional PID. Figure~\ref{fig:CERN_Setup} illustrates the CERN setup. High purity trigger signals corresponding to charged particle species were selected with information from the calorimeter and used in the setup as trigger flags. Trigger signal contamination of the other species for each particle type was below $10^{-4}$~\cite{TRT_Timepix}.
Multiple radiator configurations were tested and are shown in Table~\ref{tab:rad_layouts}; for fleece material, both 15\,cm and 23\,cm lengths were used for the GEM-TRD\_v2 and the Micromegas+GEM-TRD. Regarding Mylar foil, 15\,cm of 30\,$\upmu$m-thick foil with 240\,$\upmu$m spacing was used on the GEM-TRD, while for the Micromegas+GEM-TRD 10\,cm of 30\,$\upmu$m-thick foil with 200\,$\upmu$m spacing combined with 10\,cm of 25\,$\upmu$m-thick foil with 200\,$\upmu$m spacing for a total length of $\sim$21\,cm was used.

\subsection{Data Selection}

For in-beam measurements, several observables are recorded using fADC-125 electronics: pulse amplitude (corresponding to deposited energy) and drift time of clusters across the drift gap. Event-by-event analyses were performed to study the detector response for electrons and hadrons, enabling determination of e/$\pi$ separation power. Figure~\ref{fig:raw_f125} shows an example of raw waveforms collected for electrons in both the GEM-TRD and the Micromegas+GEM-TRD. For analysis of in-beam measurements, tight selections are applied to external PID detectors to assure sufficiently high sample purity of charged particle types. For each of the TRD prototypes tested, other detectors in the setup are used as an external track reference. Tight selections are then applied to achieve a clean sample of charged-track events and ensure maximum detection efficiency of relevant charged particle species. The detection efficiency for each prototype is calculated by reconstructing the track along one axis using external detectors and looking for a hit in the prototype that matches the external track projection into the prototype’s active area. For the signal-detection threshold used in the analysis, a typical threshold of 100\,fADC units is used where 1 fADC unit corresponds to about 0.04\,fC, or $\sim$250 electrons.

\section{Results}
\label{results}

The primary performance parameters studied were: (i) operational stability under different high-voltage configurations, (ii) response to TR photons versus ionization along the track, (iii) timing characteristics of electron and pion signals between amplification technology choices, and (iv) e/$\pi$ separation quantified by suppression factors.

At FTBF, systematic high-voltage scans were carried out for both the Micromegas-TRD and $\upmu$RWELL-TRD to map out operational stability. For the Micromegas-TRD, the drift field was scanned from about 1.3 to 1.5\,kV/cm and the mesh voltage increased to a maximum of 675\,V. Even at the maximum applied amplification voltage, the maximum detector efficiency reached was below 80\% and the detector also suffered from a decrease in operational stability due to an increase in frequency of discharges. For the $\upmu$RWELL-TRD, the drift field was scanned up to 1.6\,kV/cm, while the Micro-Well bias was increased to a maximum value of 540\,V. Similar to the Micromegas-based prototype, operation at lower amplification voltage yielded stable signals, whereas attempts to raise the voltage above $\sim$520\,V produced frequent discharges with a detection efficiency below 70\%.

\begin{figure}[!h]
\begin{center}
\includegraphics[width=0.97\columnwidth,trim={5pt 20pt 20pt 5pt},clip]{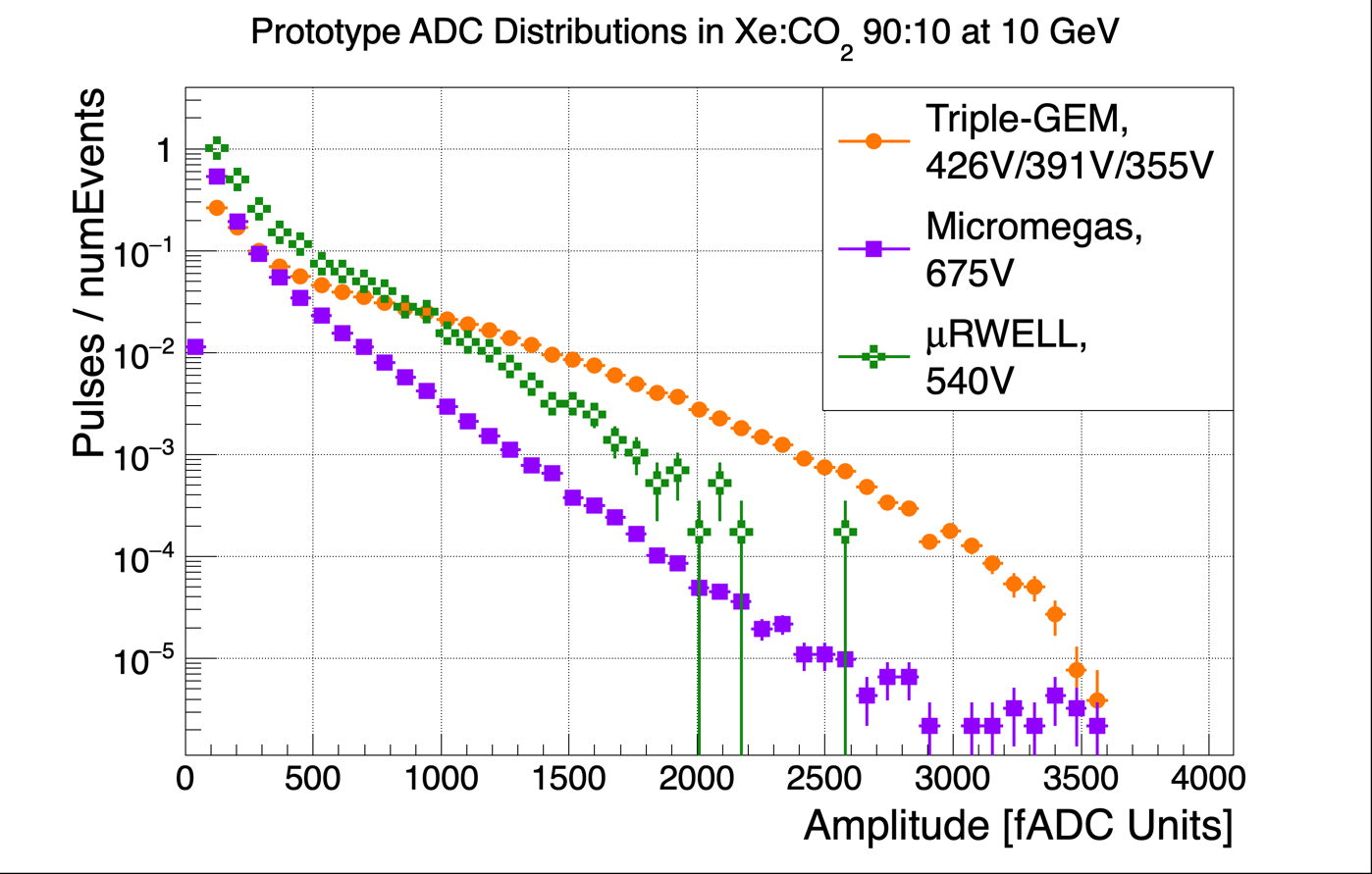}
\includegraphics[width=0.97\columnwidth,trim={5pt 20pt 20pt 5pt},clip]{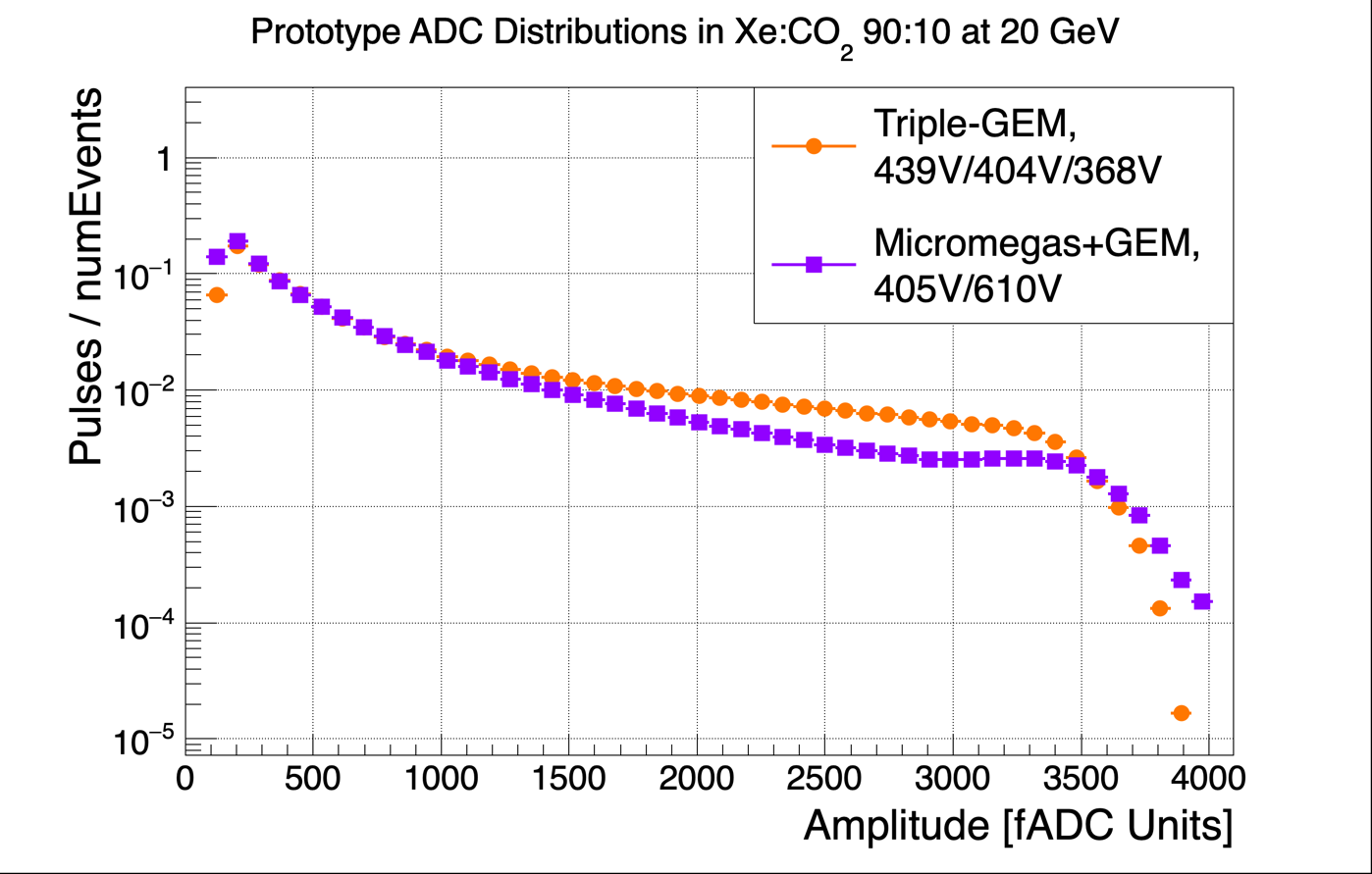}
\caption{Differential spectra for electron pulse amplitudes, normalized to the number of selected events. Each entry corresponds to an individual fADC pulse peak exceeding the fADC threshold after pedestal subtraction. (\textit{Top}) comparison of the three prototypes tested at FTBF with their maximum amplification HV values applied; (\textit{Bottom}) comparison of GEM-TRD\_v2 and Micromegas+GEM-TRD tested at CERN. The reduced gain of the Micromegas-TRD and $\upmu$RWELL-TRD prototypes is evident, as is the improved amplitude spectrum collected by the Micromegas+GEM-TRD prototype after the addition of a GEM preamplification layer. Note that voltage values are formatted in order of top-down amplification layer HV, and 1 fADC unit corresponds to $\sim$0.04\,fC.}
\label{fig:adc_spectra}
\end{center}
\end{figure}

Despite visible TR photon clusters in both the Micromegas-TRD and $\upmu$RWELL-TRD, their signal amplitude was not sufficient to reach full efficiency and extract meaningful pion suppression factors. In contrast, the GEM-TRD\_v1 exhibited stable operation across the full test campaign and served as the performance benchmark, achieving a pion detection efficiency measured to be 96.2\% with an overall statistical uncertainty of 1.9\%. Figure~\ref{fig:adc_spectra} (top) shows pulse amplitude spectra comparisons between the three prototypes at FTBF. Note that unlike Figure~\ref{fig:raw_f125}, which displays raw waveform sample distributions, Figure~\ref{fig:adc_spectra} shows differential spectra of pedestal-subtracted pulse peaks normalized to the number of selected events. Because the Micromegas prototype registered lower amplitudes than those of the GEM-TRD even at the highest amplification voltage application, a GEM preamplification layer was introduced prior to the CERN tests. At CERN, the Micromegas+GEM prototype showed improved gain and better agreement with the GEM-TRD\_v2 reference technology  (Figure~\ref{fig:adc_spectra} (bottom)). Both prototypes reached a detector efficiency plateau for all relevant charged particle species.

\begin{figure}[!h]
\begin{center}
\includegraphics[width=0.97\columnwidth,trim={5pt 5pt 20pt 5pt},clip]{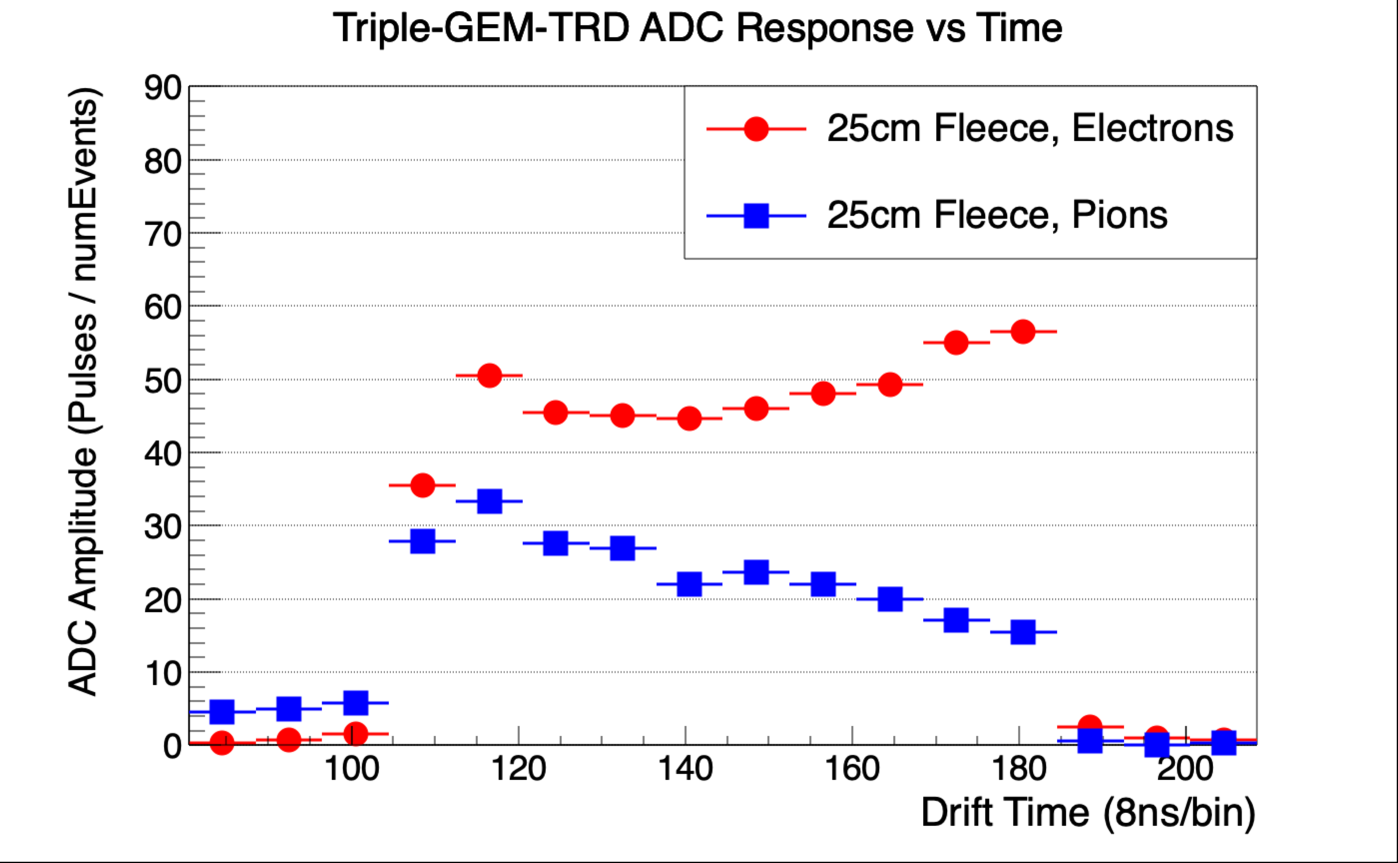}
\includegraphics[width=0.97\columnwidth,trim={5pt 7pt 20pt 5pt},clip]{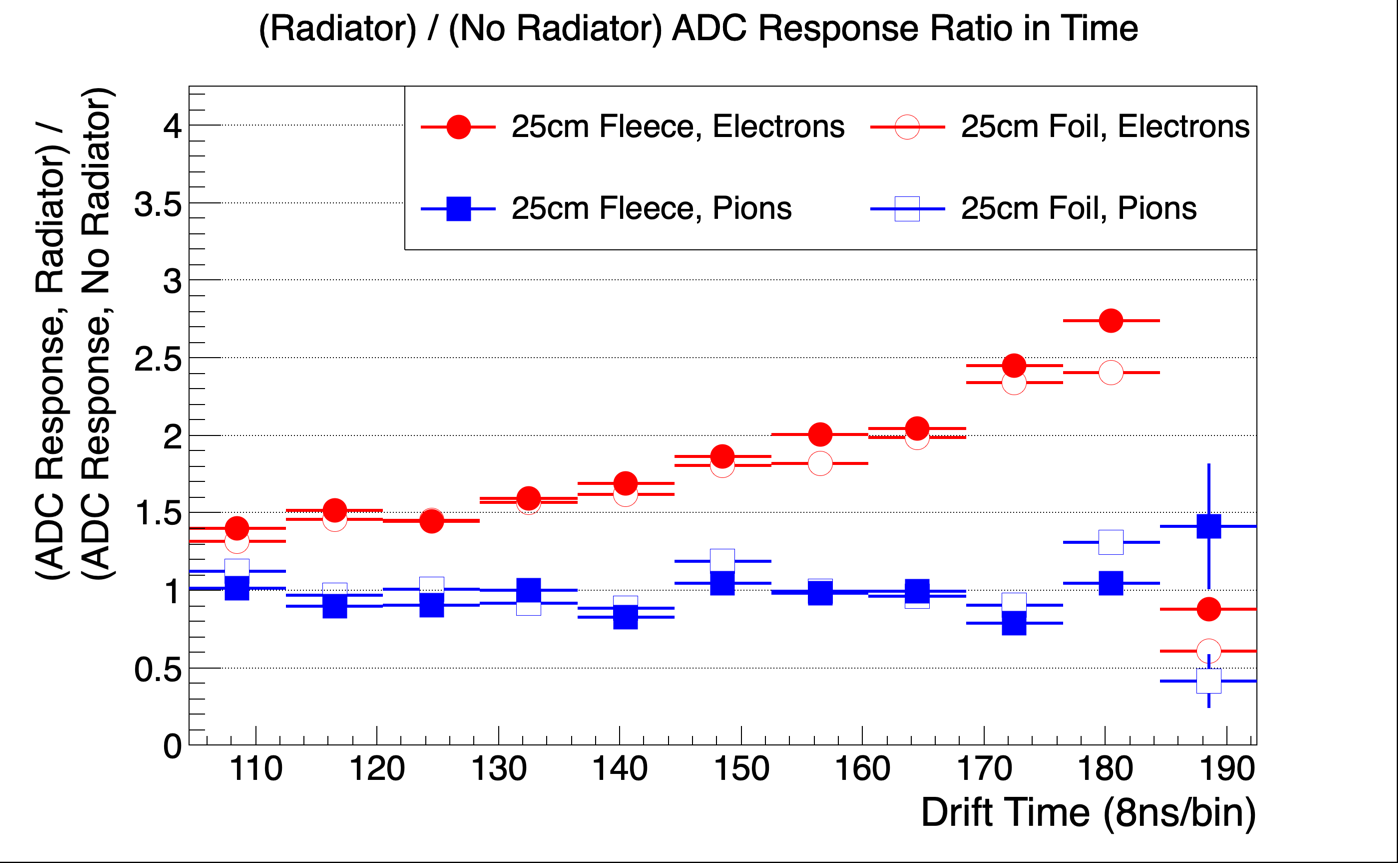}
\caption{(\textit{Top}) GEM-TRD\_v1 average ADC amplitude versus drift time with radiator at FTBF; (\textit{Bottom}) radiator versus no-radiator ratios of these averages for electrons and pions, for both fleece and foil radiators.}
\label{fig:gem_fermi_timing}
\end{center}
\end{figure}

\begin{figure}[!h]
\begin{center}
\includegraphics[width=0.99\columnwidth,trim={5pt 20pt 20pt 5pt},clip]{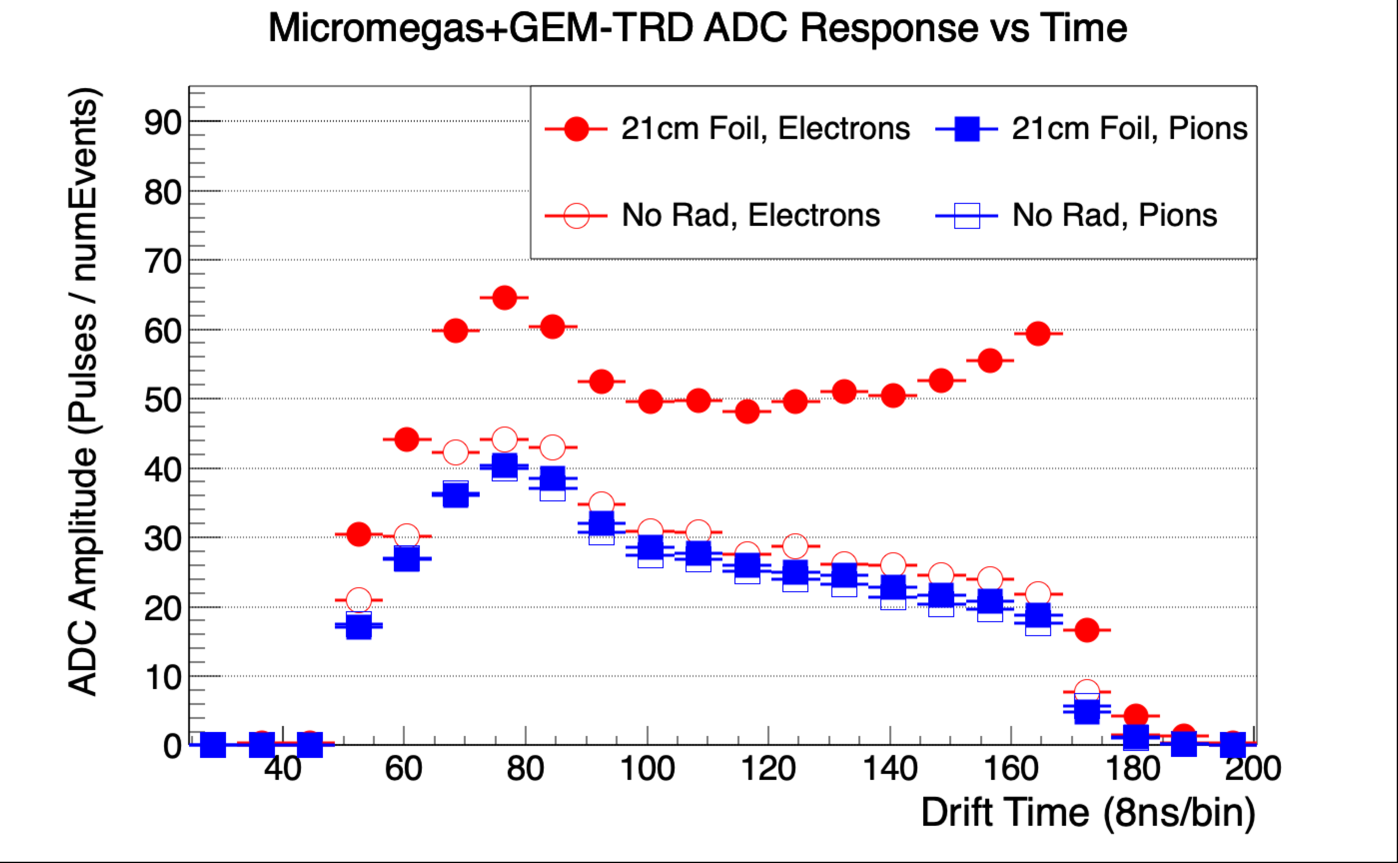}
\includegraphics[width=0.99\columnwidth,trim={5pt 20pt 20pt 5pt},clip]{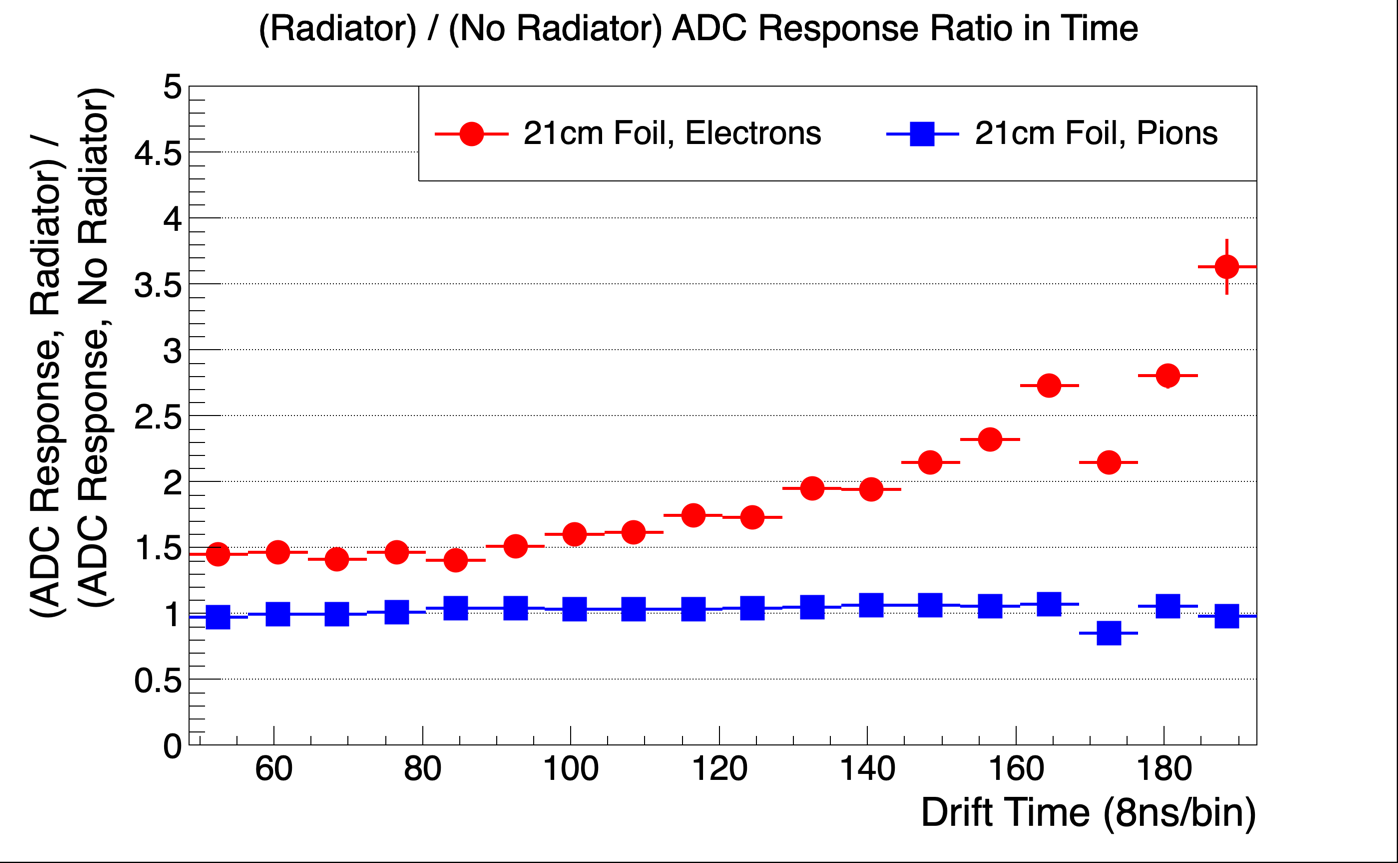}
\caption{(\textit{Top}) Micromegas+GEM-TRD average ADC amplitude versus drift time at CERN SPS; (\textit{Bottom}) radiator versus no-radiator ratios of these averages for electrons and pions.}
\label{fig:mmg_cern_timing}
\end{center}
\end{figure}

Drift time distributions were studied to distinguish TR clusters from the energy deposition along the track. Figure~\ref{fig:gem_fermi_timing} shows the GEM-TRD\_v1 drift-time response for 10\,GeV electrons and pions with radiator material present. Ratios of radiator-to-no-radiator signal shapes in time demonstrate clear excess electron signals, consistent with TR photon absorption predominantly near the cathode entrance --- equivalent to energy deposits later in drift time. The drift time distributions of the Micromegas+GEM-TRD for 20\,GeV electrons and pions are shown in Figure~\ref{fig:mmg_cern_timing}. Compared to the GEM-TRD\_v2, the overall signal extends over a longer drift time, due in part to the larger drift region of the Micromegas+GEM-TRD design and the lower drift field strength. In addition, the leading and falling edge of the prototype's distribution is less sharply defined than in the GEM-based prototype's case; this behavior is expected, as the amplification structure of a Micromegas mesh introduces slower signal collection due to the ion charge contribution~\cite{MMG} relative to the fast induction electron signals that are characteristic of GEMs. In general, the addition of the GEM preamplification layer improved timing resolution and TR photon visibility relative to the Micromegas-TRD prototype's performance at FTBF.

Pion suppression factors were derived using neural network (NN) classifiers trained on amplitude and timing observables (ROOT-based TMVA~\cite{root_tmlp}). 
Ionization along each track was used as a neural network input layer, with the particle track drift time subdivided into several slices (sum of fADC samples). Cluster counting and characterization was also used as neural network input. A multilayer perceptron with two hidden layers was implemented for suppression factor determination. The data was split into two groups: one used for training, and another (independent sample) used for final decision evaluation. Figure~\ref{fig:nn}, discussed in the next section, shows an example of the neural network output.

\begin{figure}[!h]
\begin{center}
\includegraphics[width=0.97\columnwidth,trim={5pt 5pt 20pt 5pt},clip]{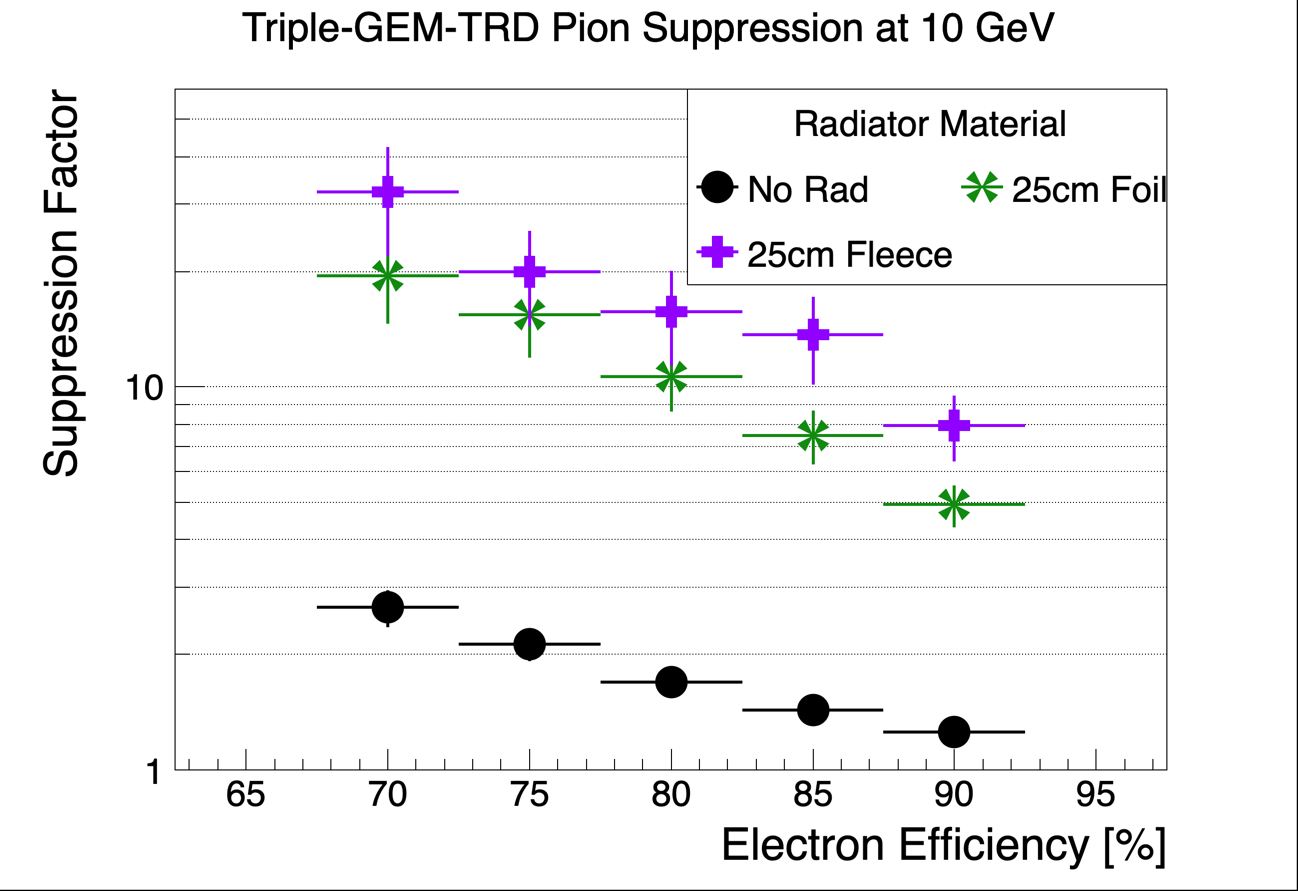}
\caption{Pion suppression factors for the GEM-TRD\_v1 for various electron NN efficiencies used in 10\,GeV beam. Statistical error bars are dominated by the pion sample statistics, as visualized by the example in Figure~\ref{fig:nn}. These values have an overall systematic error of 1.9\% resulting from the pion efficiency uncertainty.}
\label{fig:fermi_rej}
\end{center}
\end{figure}

\begin{figure}[!h]
\begin{center}
\includegraphics[width=0.97\columnwidth,trim={5pt 4pt 20pt 5pt},clip]{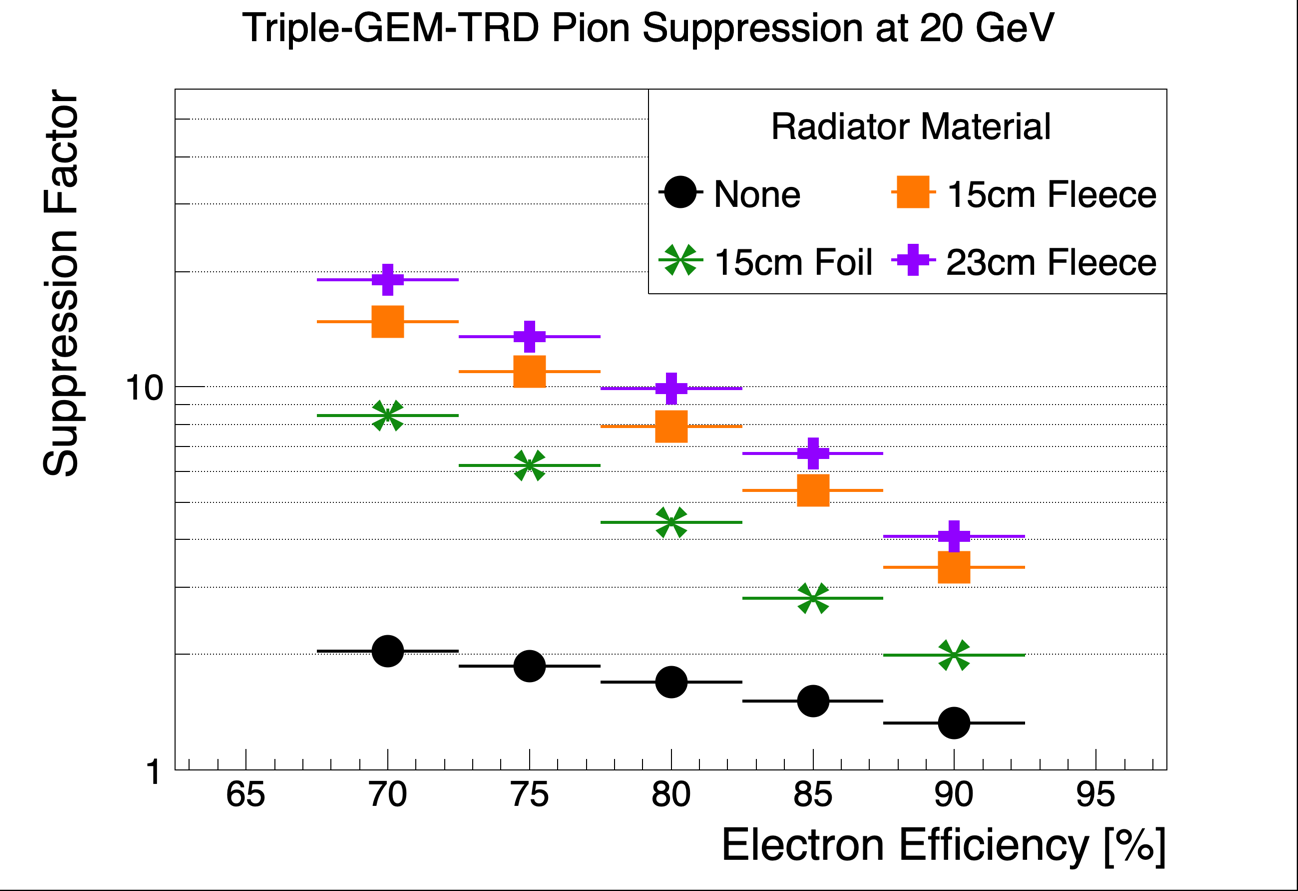}
\includegraphics[width=0.97\columnwidth,trim={5pt 4pt 20pt 5pt},clip]{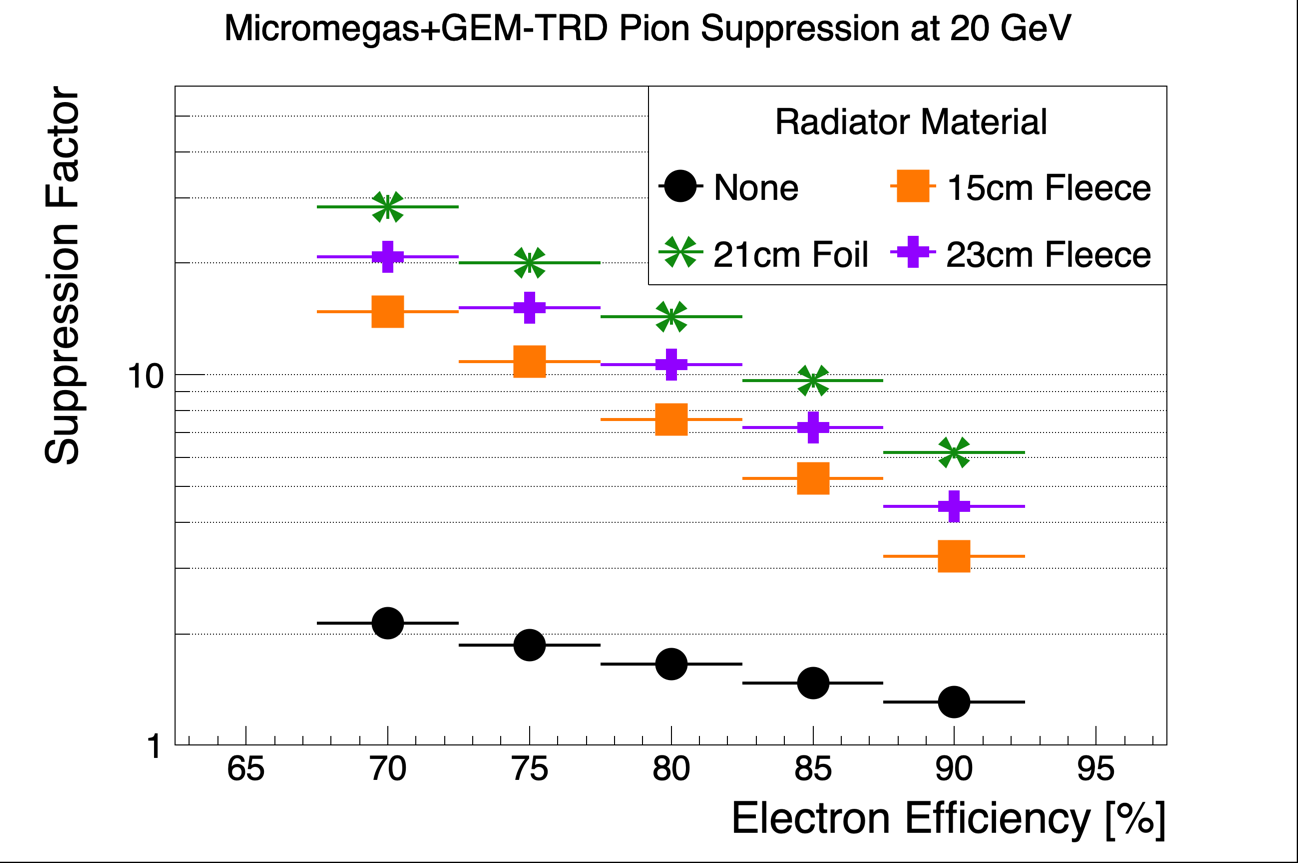}
\caption{Pion suppression factors for the GEM-TRD\_v2 (\textit{Top}) and for the Micromegas+GEM-TRD (\textit{Bottom}) for various electron NN efficiencies used in 20\,GeV beam. Statistical error bars are dominated by the pion sample statistics, as visualized by the example in Figure~\ref{fig:nn}.}
\label{fig:cern_rej}
\end{center}
\end{figure}

At FTBF, the GEM-TRD\_v1 achieved a pion suppression factor around 8 at 10\,GeV for a representative working point of 90\% electron efficiency with the fleece radiator (Figure~\ref{fig:fermi_rej}). It is important to note that this efficiency value is not a detector limitation, but rather one of several benchmark points used to quantify performance as depicted in Figure~\ref{fig:fermi_rej}. No reliable suppression factors were extracted for the Micromegas-TRD or the $\upmu$RWELL-TRD at FTBF due to gain limitations previously discussed. At CERN, both the GEM-TRD\_v2 and Micromegas+GEM-TRD were evaluated - Figure~\ref{fig:cern_rej} shows the suppression factors for each detector obtained at 20\,GeV for different radiator options.
Note that the significant difference in each prototype's performance with the foil-based radiator is explained by the difference in overall radiator length and construction as described in Section~\ref{cern_setup} and Table~\ref{tab:rad_layouts}, with the foil-based radiator used on the Micromegas+GEM-TRD being better optimized for TR generation and transmission.

The Micromegas+GEM-TRD generally showed comparable pion suppression performance to the GEM-TRD\_v2, and the authors note that the difference in suppression between the two prototypes is partly attributable to the larger drift gap in the Micromegas-based prototype design compared to the GEM-TRD's. Neither detector was able to reach the suppression performance of the GEM-TRD\_v1 seen at FTBF.

\section{Discussion}
\label{discussion}

\begin{figure}[!h]
\begin{center}
\includegraphics[width=0.97\columnwidth,trim={80pt 0pt 35pt 30pt},clip]{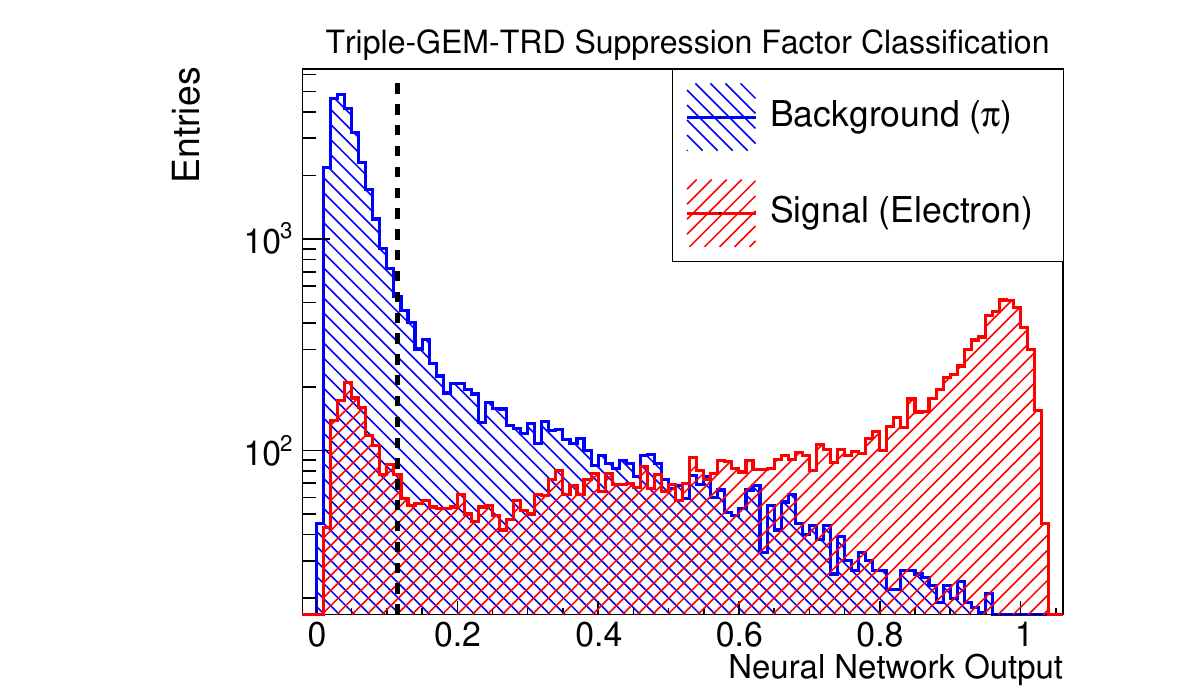}
\caption{Example of NN output for the GEM-TRD\_v2 in 20\,GeV mixed electron--hadron beam. The dashed vertical line is plotted at the 90\% electron efficiency value as an example of sample purity selection. Note that for the suppression factor determination, the pion tail to the right of the plotted selection is what dominates the statistical error.}
\label{fig:nn}
\end{center}
\end{figure}

Previous Monte Carlo simulations discussed in~\cite{NIMGEMTRD1} predict a pion suppression factor of roughly 8 for a reference triple-GEM-TRD with 15\,cm of radiator in Xe:CO$_{2}$ 90:10. The suppression performance of the GEM-TRD\_v1 in 10\,GeV mixed beam resulted in a suppression factor of about 8 with 25\,cm of fleece radiator, while the CERN beam tests at 20\,GeV yielded suppression factors of $\sim$4 for the GEM-TRD\_v2 and $\sim$4.5 for the Micromegas+GEM-TRD at 90\% electron efficiency with similar radiator settings. The authors attribute this significant discrepancy between previous results and these measurements primarily to the change in cathode material. Whereas the GEM-TRD\_v1 prototype utilized a 0.2\,$\upmu$m chromium cathode, both prototypes tested at CERN SPS employed a 5\,$\upmu$m copper cathode. The TR photon spectrum relevant for absorption in Xe:CO$_{2}$ spans roughly 3--40\,keV, peaking near 11\,keV. Copper has a strong K-$\alpha$ absorption edge near 9\,keV~\cite{xray}, meaning that the copper cathode absorbs a significant fraction of the TR energy spectrum exiting the radiator before it has a chance to enter the detector's gas volume. This material choice would naturally reduce the number of detectable TR photons and therefore degrade the pion suppression factor. Chromium also has a K-$\alpha$ absorption line near 6\,keV, causing a similar though less significant effect in the GEM-TRD\_v1 performance. Figure~\ref{fig:nn} contains an example of the NN output for the GEM-TRD\_v2 based on electron and hadron signals collected at CERN SPS. The enhancement in the signal to the left of the 90\% efficiency line is presumably a result of two separate occurrences: electrons which pass through the prototype without emission of TR photons, and electrons whose TR photon emission are not absorbed in the gaseous detector region.

\begin{figure*}[!h]
\begin{center}
\includegraphics[width=0.65\textwidth,trim={16pt 29pt 23pt 25pt},clip]{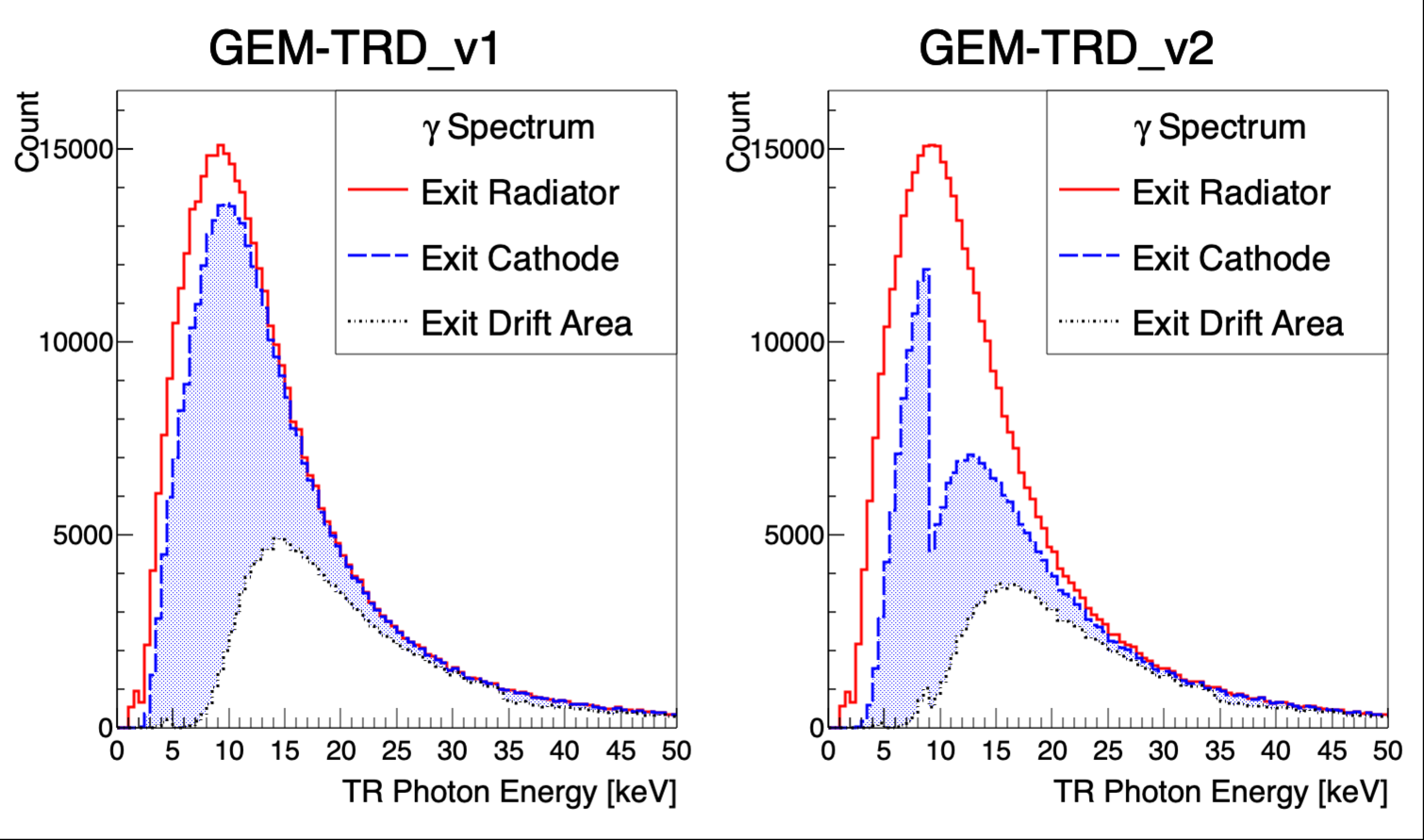}
\includegraphics[width=0.65\textwidth,trim={16pt 29pt 23pt 25pt},clip]{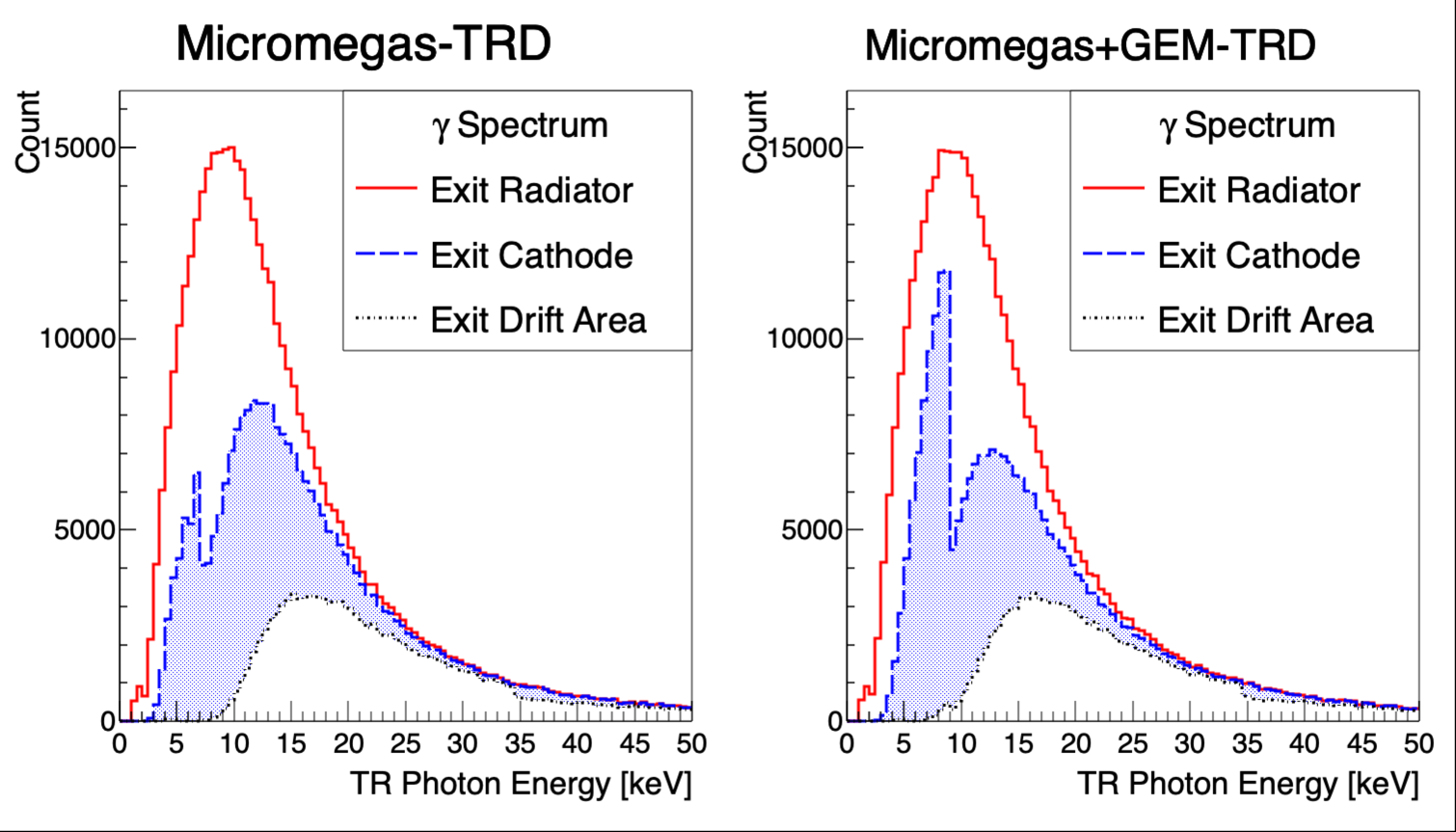}
\caption{\texttt{Geant4} Simulation of the energy spectra of TR photons generated from 15\,cm of regular foil radiator, with the shaded region representing the spectrum that experiences absorption in the gaseous drift region of the tested prototypes. (\textit{Top Left}) The GEM-TRD\_v1 construction with spectra generated by 10\,GeV electrons; (\textit{Top Right}) the GEM-TRD\_v2 construction with spectra from 20\,GeV electrons. (\textit{Bottom Left}) The Micromegas-TRD design with spectra from 10\,GeV electrons; (\textit{Bottom Right}) the Micromegas+GEM-TRD design with spectra from 20\,GeV electrons.}
\label{fig:geant_compare}
\end{center}
\end{figure*}

Monte Carlo studies using the \texttt{Geant4} TR package~\cite{geant4_trd} were performed to model radiator photon yields and absorption in sequential layers of the detector prototypes. Figure~\ref{fig:geant_compare} shows predicted TR photon spectra for 15\,cm of regular-spaced radiator consisting of polypropylene foils of the same consistency as those used in the FTBF tests. Top plots in Figure~\ref{fig:geant_compare} show results for the different GEM-TRD prototype constructions that were tested, for 10\,GeV and 20\,GeV electrons passing through, respectively. Bottom plots show results for 10\,GeV and 20\,GeV electrons passing through the Micromegas-TRD design and the Micromegas+GEM-TRD construction.

These results emphasize that TRD optimization requires careful balancing of multiple design factors: radiator thickness and material for sufficient TR yield, cathode material composition and thickness for minimal soft TR X-ray stoppage, drift gap size for photon absorption depth, and amplification technology for signal timing and robustness. This work stands to establish clear design requirements for MPGD-based TRD operation and demonstrates that multi-stage amplification is not merely advantageous but necessary for achieving meaningful TRD performance with MPGDs. As a followup study, the authors have tested a variety of cathode materials on each prototype at Jefferson Lab in 3--6\,GeV electron beam in order to directly demonstrate and quantify this effect. This ongoing R\&D effort will also focus on refining these multi-stage-gain detector configurations to maximize suppression performance while minimizing material and complexity, guiding the development of next-generation TRDs for high-rate experimental environments. Analysis of these studies is currently underway and will be reported in forthcoming publications.

\section{Summary and Future Work}
\label{conclusion}

We have reported on prototype development and beam test results for two triple-GEM-TRDs, a Micromegas-TRD, a Micromegas+GEM-TRD, and a $\upmu$RWELL-TRD exposed to electron and pion beams at FTBF (10\,GeV) and CERN SPS (20\,GeV). The GEM-TRD\_v1 achieved a pion suppression factor of about 8 at 90\% electron efficiency with a 25\,cm fleece radiator. Both the GEM-TRD\_v2 and the Micromegas+GEM-TRD prototypes yielded significantly lower suppression factors - approximately 4 and 4.5, respectively, with a 23\,cm fleece radiator. The principal factor identified is the cathode material, providing a natural explanation for the degraded suppression performance observed. Prototypes tested in 20\,GeV beam utilized a 5\,$\upmu$m copper cathode, which eminently absorbs TR photons in the relevant energy range, thereby reducing the number of photons reaching the Xe:CO$_{2}$ absorption region. Followup tests at Jefferson Lab have systematically varied cathode composition to quantify this impact and establish conditions under which suppression factors comparable to the GEM-TRD\_v1 design can be determined.

Drift time measurements further underscore the importance of amplification structure and drift configuration in a TRD application. The Micromegas-based prototype exhibited longer collection times and less sharply defined leading edges than the GEM-based prototypes, consistent with its larger drift gap and slower ion-dominated signal formation. These observations highlight an additional design trade-off: maximizing TR photon absorption through larger drift regions must be balanced against timing resolution for applications where charged particle tracking is a substantial component.

To summarize the findings resulting from these studies:
\begin{itemize}
    \item The GEM-TRD remains the most mature technology, achieving stable operation and a pion suppression factor up to $\sim$8 at 90\% electron efficiency in mixed beams with presumably improved performance following design changes outlined in Section~\ref{discussion}.
    \item The Micromegas-based TRD, with the addition of a GEM preamplification stage, showed clear improvement in operational stability and gain at CERN SPS relative to FTBF and demonstrated clear absorption and discrimination of TR photon signals to charged track d\textit{E}/d\textit{x}.
    \item The $\upmu$RWELL-TRD faced gain limitations and was unable to achieve suppression factor measurements, but with the addition of a GEM preamplification layer similar to the Micromegas-based prototype, may be another feasible option for a TRD application. 
    \item Radiator comparisons confirmed expectations: fleece provides broader spectra and higher total photon yield compared with foil-based radiators, the latter of which offer reduced material density.
\end{itemize}

Overall, this work represents the first in-beam measurements of Micromegas- and $\upmu$RWELL-based TRDs, and an expansion on the understanding of performance capabilities for a triple-GEM-TRD. While GEM-based design remains the most reliable choice, continued development of TRDs based on Micromegas and $\upmu$RWELL (with staged gain and optimized cathode transparency) appear likely to provide viable alternatives. Ongoing studies will refine this understanding and guide the next generation of TRD prototypes toward optimal suppression performance in the energy regime of interest. This coordinated R\&D effort demonstrates that MPGDs will serve as the foundation for a next-generation TRD.

\section*{Acknowledgments}

The authors thank the staff at the Fermilab Test Beam Facility, including Mandy Rominsky, Eugene Schmidt, Joe Pastika, and Todd Nebel for their support and expertise during the beam test campaign. We extend similar thanks to the staff at the CERN SPS H8 beamline, including Barbara Holzer and Martin Jaekel. We also thank Matt Posik and Bernd Surrow of Temple University for technical contributions to the FTBF campaign, and the ATLAS TRT group, including Semen Doronin, Anatoli Romaniouk, Konstanin Vorobev, and Konstantin Zhukov for their expertise and involvement during the CERN SPS campaign. This material is based upon work supported by the U.S. Department of Energy, Office of Science, Office of Nuclear Physics under contract DE-AC05-06OR23177. This work was also supported in part by Department of Energy Award DE-FG05-92ER40712, DOE EIC Generic R\&D 2022\_02, and Vanderbilt University. The work of A.A. was supported by the DOE, Office of Science, Office of Nuclear Physics in the Early Career Program. This material is based upon work supported by the U.S. Department of Energy, Office of Science, Office of Workforce Development for Teachers and Scientists, Office of Science Graduate Student Research (SCGSR) program administered by the Oak Ridge Institute for Science and Education for the DOE under contract number DE‐SC0014664. Finally, the authors thank Beni Zihlmann from Jefferson Lab and University of Virginia members Huong Nguyen and Nilanga Liyanage for their longtime support and involvement with these efforts.

\appendix

\section{TR Radiators}
\label{App_r}

See Table~\ref{tab:rad_layouts}.

\begin{table*}[ht]
    \scriptsize
    \centering
    \renewcommand{\arraystretch}{1.2}
    \setlength{\tabcolsep}{4pt}
    \begin{tabular}{lllllllll}
\hline
&\multicolumn{2}{c}{GEM-TRD\_v1}&\multicolumn{2}{c}{GEM-TRD\_v2}&\multicolumn{2}{c}{Micromegas-TRD}&\multicolumn{2}{c}{Micromegas+GEM-TRD}\\
\cline{2-3} \noalign{\vskip 2.5pt}
\cline{4-5} \noalign{\vskip -2.5pt}
\cline{6-7} \noalign{\vskip 2.5pt}
\cline{8-9} \noalign{\vskip -2.5pt}
Parameter\\
\hline
Beam energy&10\,GeV&10\,GeV&20\,GeV&20\,GeV&10\,GeV&10\,GeV&20\,GeV&20\,GeV\\
Rad. material&Mylar&Fleece&Mylar&Fleece&Mylar&Fleece&Mylar&Fleece\\
Rad. layer thickness [$\upmu$m]&12.5&2500&30&2500&12.5&2500&30\&25&2500\\
Gap material&Netted spacer&Air&Air&Air&Air&Netted spacer&Air&Air\\
Gap thickness [$\upmu$m]&300&N/A&240&N/A&300&N/A&200\&200&N/A\\
N layers&474&100&555&60, 90&474&100&450\&450&60, 90\\
Overall length [cm]&25&25&15&15, 23&25&25&21&15, 23\\
\hline
\end{tabular}
    \caption{Radiator configurations used for the different TRD prototypes tested at FTBF [10\,GeV] and CERN SPS [20\,GeV]. Descriptions are provided in the text of sections~\ref{ftbf_setup} and~\ref{cern_setup}, respectively. Note that for the fleece radiator, the material is woven polypropylene fibers of 20\,$\upmu$m diameter.}
    \label{tab:rad_layouts}
\end{table*}

\section{High Voltage Field Configurations}
\label{App_a}

See Tables~\ref{tab:gem_divider} and~\ref{tab:mmg_divider}.

\begin{table}[!h]
    \scriptsize
    \centering
    \begin{tabular}{llll}
\hline
& \multicolumn{2}{c}{Xe:CO$_{2}$ Settings} \\
\cline{2-4}
Quantity & GEM-TRD\_v1 & GEM-TRD\_v2 \\
\hline
Drift gap E-field [kV/cm] & 1.38 & 1.52 \\
GEM1 E-field [kV/cm] & 85.2 & 87.8 \\  
TG1 E-field [kV/cm] & 3.56 & 3.62 \\
GEM2 E-field [kV/cm] & 78.2 & 80.8 \\
TG2 E-field [kV/cm] & 3.56 & 3.62 \\
GEM3 E-field [kV/cm] & 71 & 73.6 \\
IG E-field [kV/cm] & 3.56 & 3.45 \\

Effective Gain & $0.8~\times~10^{4}$ & $1.4~\times~10^{4}$ &  \\
\hline
\end{tabular}
    \caption{High voltage field configurations for the GEM-TRDs for measurements taken with Xe:CO$_{2}$ 90:10. Conventionally, GEM1 is the top-most layer, GEM2 is the middle layer, and GEM3 is the bottom-most layer. `TG' denotes Transfer Gap and `IG' denotes Induction Gap.}
    \label{tab:gem_divider}
\end{table}

\begin{table}[!h]
    \scriptsize
    \centering
    \begin{tabular}{lll}
\hline
& \multicolumn{2}{c}{Xe:CO$_{2}$ Settings} \\
\cline{2-3}
Quantity &  \\
\hline
Drift gap E-field [kV/cm] & & 1.36 \\
GEM E-field [kV/cm] & & 81.5 \\  
TG E-field [kV/cm] & & 2.45  \\

Voltage on mesh [V] & & 610  \\

Effective Gain & & $1.0~\times~10^{4}$  \\
\hline
\end{tabular}
    \caption{High voltage field configuration for the Micromegas+GEM-TRD for measurements taken with Xe:CO$_{2}$ 90:10. `TG' denotes Transfer Gap.}
    \label{tab:mmg_divider}
\end{table}

\bibliographystyle{elsarticle-num} 
\bibliography{theBibliography}

\end{document}